\documentclass[12pt]{article}

\usepackage{epsf}
\renewcommand{\bar}[1]{\overline{#1}}

\begin{document}

\begin{flushright}
USM--TH--136  \\
SLAC--PUB--9677\\
September 2004
\end{flushright}

\bigskip\bigskip

\vspace{1.5in}

 \centerline{{\large \bf Nuclear Antishadowing
in Neutrino Deep Inelastic Scattering}\footnote{Work supported in
part by the Department of Energy under contract number
DE--AC03--76SF00515, by the Foundation for University Key Teacher
by the Ministry of Education (China), by Fondecyt (Chile) grant
1030355, and by the National Natural Science Foundation of China
under Grant Numbers 19875024 and 10025523. \\ }}

\vspace{22pt}

\centerline{\bf { Stanley J. Brodsky
\footnote{e-mail:sjbth@SLAC.Stanford.EDU}$^{a}$, Ivan
Schmidt\footnote{e-mail: ivan.schmidt@fis.utfsm.cl}$^{b}$,
Jian-Jun Yang\footnote{deceased}$^{b,c,d}$}}

\vspace{8pt}

{\centerline {$^{a}$Stanford Linear Accelerator Center, Stanford University,}}

{\centerline {Stanford, California 94309,}}

{\centerline {$^{b}$Departamento de F\'\i sica, Universidad
T\'ecnica Federico Santa Mar\'\i a,}}

{\centerline {Casilla 110-V,
Valpara\'\i so, Chile}}

{\centerline {$^{c}$Institut f\"{u}r Theoretische Physik,
Universit\"{a}t Regensburg,}

{\centerline  {D-93040 Regensburg, Germany,}}

{\centerline {$^{d}$Department of Physics, Nanjing Normal
University,}}

{\centerline {Nanjing 210097, China}}

\vfill \newpage

 {\hbox{\vspace{1in}}

\begin{center}
{\large \bf Abstract}
\end{center}

The shadowing and antishadowing of nuclear structure functions in
the Gribov-Glauber picture is due respectively to the destructive
and constructive interference of amplitudes arising from the
multiple-scattering of quarks in the nucleus.   The effective
quark-nucleon scattering amplitude includes Pomeron and Odderon
contributions from multi-gluon exchange as well as Reggeon
quark-exchange contributions.  We show that the coherence of these
multiscattering nuclear processes leads to shadowing and
antishadowing of the electromagnetic nuclear structure functions
in agreement with measurements.  This picture leads to
substantially different antishadowing for charged and neutral
current reactions, thus affecting the extraction of the
weak-mixing angle $\theta_W$. We find that part of the anomalous
NuTeV result for $\theta_W$ could be due to the nonuniversality
of nuclear antishadowing for charged and neutral currents.
Detailed measurements of the nuclear dependence of individual
quark structure functions are thus needed to establish the
distinctive phenomenology of shadowing and antishadowing and to
make the NuTeV results definitive.

\newpage

\section{Introduction}

The precise determination of the weak-mixing angle $\sin^2\theta_W$ plays a crucial role
in testing the standard model of electroweak interactions.  Until recently, a consistent
value was obtained from all of the electroweak observables~\cite{Abbaneo:2001ix}.
However, the NuTeV Collaboration~\cite{Zeller:2001hh} has determined a value for
$\sin^2\theta_W$ from measurements of the ratio of charged and neutral current deep
inelastic neutrino--nucleus and anti-neutrino--nucleus scattering in iron targets which
has a $3 \sigma$ deviation with respect to the fit of the standard model predictions from
other electroweak measurements~\cite{Abbaneo:2001ix}.  This contrasts with the recent
determination of $\sin^2\theta_W$ from parity violation in M\"oller scattering which is
consistent with the standard model~\cite{Anthony:2003ub}.  Although the NuTeV analysis
takes into account many sources of systematic errors, there still remains the question of
whether the reported deviation could be accounted for by QCD effects such as the
asymmetry of the strange-antistrange quark sea~\cite{Burkardt:1991di,Brodsky:1996hc} or
other Standard Model
effects~\cite{KSY,Miller:2002xh,Zeller:2002et,Zeller:2002du,McFarland:2003gx,
Diener:2003ss,Kulagin:2003wz,Bernstein:2002sa,Kretzer:2004tv,
Kretzer:2003wy,Davidson:2001ji}. In this paper we shall investigate whether the anomalous
NuTeV result for $\sin^2\theta_W$ could be due to the different behavior of leading-twist
nuclear shadowing and antishadowing effects for charged and neutral currents.

The physics of the nuclear shadowing in deep inelastic scattering
can be most easily understood in the laboratory frame using the
Glauber-Gribov picture~\cite{Glauber:1955qq,Gribov:1968gs}.  The
virtual photon, $W$ or $Z^0$  produces a quark-antiquark
color-dipole pair which can interact diffractively or
inelastically on the nucleons in the nucleus. The destructive
interference of diffractive amplitudes from pomeron exchange on
the upstream nucleons then causes shadowing of the virtual photon
interactions on the back-face
nucleons~\cite{Stodolsky:1966am,Brodsky:1969iz,Ioffe:1969kf,
Frankfurt:1988zg,Kopeliovich:1998gv,Kharzeev:2002fm}. As
emphasized by Ioffe~\cite{Ioffe:1969kf}, the coherence between
processes which occur on different nucleons at separation $L_A$
requires small Bjorken $x_{B}:$ $1/M x_B = {2\nu/ Q^2}  \ge L_A .$
The coherence between different quark processes is also the basis
of saturation phenomena in DIS and other hard QCD reactions at
small $x_B$~\cite{Mueller:2004se}, and coherent multiple parton scattering
has been used in the analysis of $p+A$ collisions in terms of the perturbative
QCD factorization approach~\cite{Qiu:2004da}. An example of the interference
of one- and two-step processes in deep inelastic lepton-nucleus
scattering illustrated in Fig.~\ref{bsy1f1}.

\vspace{0.3cm}
\begin{figure}[htb]
\begin{center}
\leavevmode {\epsfysize=10cm \epsffile{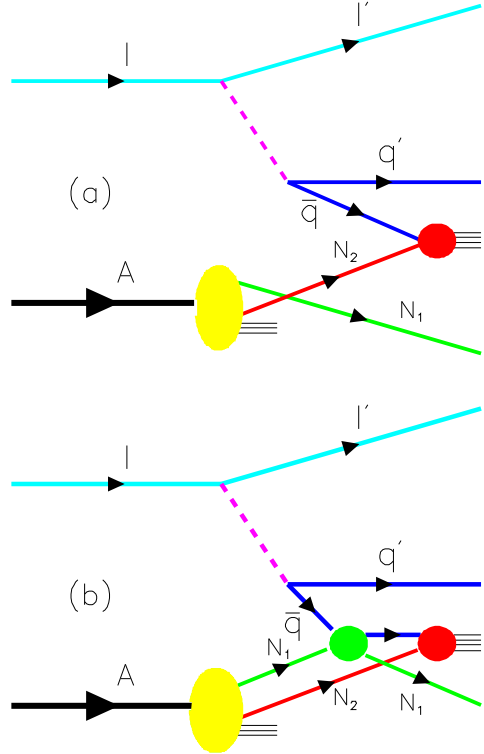}}
\end{center}
\caption[*]{\baselineskip 13pt The one-step and two-step processes
in DIS on a nucleus.  If the scattering on nucleon $N_1$ is via
pomeron exchange, the one-step and two-step amplitudes are
opposite in phase, thus diminishing the $\bar q$ flux reaching
$N_2.$ This causes shadowing of the charged and neutral current
nuclear structure functions. \label{bsy1f1}}
\end{figure}

An important aspect of the shadowing phenomenon is that the
diffractive contribution $\gamma^* N \to X N^\prime$ to deep
inelastic scattering (DDIS) where the nucleon $N_1$ in
Fig.~\ref{bsy1f1} remains intact is a constant fraction of the
total DIS rate, confirming that it is a leading-twist
contribution. The Bjorken scaling of DDIS has been observed at
HERA~\cite{Martin:2004xw,Adloff:1997sc,Ruspa:2004jb}. As shown in
Ref.~\cite{Brodsky:2002ue}, the leading-twist contribution to DDIS
arises in QCD in the usual parton model frame when one includes
the nearly instantaneous gluon exchange final-state interactions
of the struck quark with the target spectators. The same final
state interactions also lead to leading-twist single-spin
asymmetries in semi-inclusive DIS~\cite{Brodsky:2002cx}. Thus the
shadowing of nuclear structure functions is also a leading-twist
effect.

It was shown in Ref.~\cite{BHPRL90}  that if one allows for
Reggeon exchanges which leave a nucleon intact,  then one can
obtain {\it constructive} interference among the multi-scattering
amplitudes in the nucleus.   A Bjorken-scaling contribution to
DDIS from Reggeon exchange has in fact also been observed at
HERA~\cite{Adloff:1997sc,Ruspa:2004jb}. The strength and energy
dependence of the $C=+$ Reggeon $t-$channel exchange contributions
to virtual Compton scattering is constrained by the
Kuti-Weisskopf~\cite{Kuti:1971ph} behavior $F_2(x) \sim
x^{1-\alpha_R}$ of the non-singlet electromagnetic structure
functions at small $x$.  The phase of the Reggeon exchange
amplitude is determined by its signature factor.  Because of this
complex phase structure~\cite{BHPRL90}, one obtains constructive
interference and {\it antishadowing} of the nuclear structure
functions in the range $0.1 < x < 0.2$ -- a pronounced excess of
the nuclear cross section with respect to nucleon
additivity~\cite{Arneodo:1992wf}.

In the case where the diffractive amplitude on $N_1$ is imaginary,
the two-step process has the phase $i \times i = -1 $ relative to
the one-step amplitude, producing destructive interference. (The
second factor of $i$ arises from integration over the quasi-real
intermediate state.)  In the case where the diffractive amplitude
on $N_1$ is due to $C=+$ Reggeon exchange with intercept
$\alpha_R(0) = 1/2$, for example, the phase of the two-step
amplitude is ${1\over \sqrt 2}(1-i) \times i = {1\over \sqrt 2}
(i+1)$ relative to the one-step amplitude, thus producing
constructive interference and antishadowing. This is discussed in
more detail in the following sections.

Odderon exchange due to three-gluon exchange leads to an elastic
quark-nucleon amplitude which is nearly real in phase, thus
providing an additional mechanism for antishadowing. We shall show
that the combination of Pomeron, Reggeon, and Odderon exchanges in
multi-step processes leads to shadowing and antishadowing of the
electromagnetic nuclear structure functions in agreement with
measurement in electromagnetic interactions. Shadowing of the
nuclear structure functions is thus due to the dynamics of
$\gamma^* A$ interactions; it is not a property of the nuclear
light-front wavefunction computed in
isolation~\cite{Brodsky:2002ue}.

Evidence for the Odderon has been illusive; a detailed discussion
can be found in a recent review by Ewerz~\cite{EWERZ03}.  A clear
signal appears in the difference of proton-proton vs. proton
anti-proton scattering.  From the perspective of QCD, the Odderon
represents the color-singlet effects of three gluons in the
$t-$channel.  A general treatment in the context of the BFKL
program has been given by Bartels, Lipatov, and
Vacca~\cite{Bartels:1999yt}  The Odderon has Regge intercept
$\alpha_{\cal O} \sim 1$, $C=-,$ and thus its phase is nearly pure
real.  The Odderon does not contribute directly to the structure
functions since it gives a real contribution to the virtual
Compton amplitude.  However, it can play an important role in the
multi-scattering series in the nuclear target.

There can be other important antishadowing mechanisms.  Processes
which can occur on a nucleus, but are forbidden on a nucleon, will
enhance the nuclear structure functions. For example, pseudoscalar
Reggeon exchange amplitudes do not contribute to DIS on a nucleon
target since the helicity-conserving forward amplitude $\gamma^* N
\to \gamma^* N$ vanishes at $t=0$.  However in the nuclear case,
the interactions of the scattered quark (due to pomeron exchange)
on a second nucleon $N_2$ in a nuclear target can skew the
kinematics of $\gamma^* N_1 \to \gamma^* N_1^\prime,$ thus
allowing the pseudoscalar exchange to occur on the nucleon $N_1$
at $t \ne 0$.  This also requires nonzero orbital angular momentum
of the nucleons in the nuclear wavefunction. Notice that the
virtual Compton amplitude on the nucleus $\gamma^* A \to \gamma^*
A$ is still evaluated at zero momentum transfer $t=0$. Thus in
general one should include pseudoscalar exchange in the
parametrization of the quark multiple scattering processes.

\vspace{0.3cm}
\begin{figure}[htbp]
\begin{center}
\leavevmode {\epsfysize=10cm \epsffile{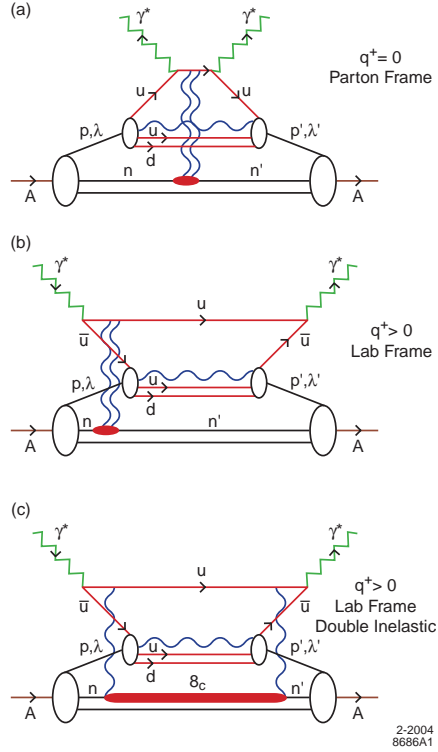}}
\end{center}
\caption[*]{\baselineskip 13pt Representation of leading-twist QCD
contributions to the nuclear structure function as calculated from
the absorptive part of the forward virtual Compton amplitude {\it
Im}$~ T(\gamma^* A \to \gamma^* A).$ (a) Illustration of a
two-step contribution in the usual $q^+ = 0, q^2_\perp = Q^2$
parton model frame. The deep inelastic scattering of a lepton on a
valence quark of a target proton is followed by the final-state
two-gluon exchange ``pomeron" interaction of the outgoing quark on
a neutron. (b) Illustration of the physics of the same two-step
process illustrated in (a), but in the laboratory frame where $q^+
> 0$.  The $u \bar u$ fluctuation of the virtual photon scatters
elastically via two-gluon exchange on a neutron; this is then
followed by the annihilation of the $\bar u$ quark on a proton.
(c) Illustration of a doubly inelastic discontinuity of the same
two-step process as (a) in the laboratory frame $q^+ > 0$.  The $u
\bar u$ fluctuation of the virtual photon first scatters
inelastically on a neutron via a single-gluon exchange which
produces an excited color state $8_C$ of the neutron.  This is
then followed by the annihilation of the $\bar u$ quark on a
proton.

\label{8686A01}}
\end{figure}

In Fig.~\ref{8686A01} we illustrate several leading-twist QCD
contributions to the nuclear structure function as calculated from
the absorptive part of the forward virtual Compton amplitude {\it
Im}$~ T(\gamma^* A \to \gamma^* A),$ in the $q^+ = 0, q^2_\perp =
Q^2$ parton model frame and in the laboratory frame where $q^+ >
0$.  Notice the final-state two-gluon exchange ``pomeron"
interaction of the outgoing quark on a target neutron.
Figure~\ref{8686A01} (c) is an illustration of a doubly inelastic
discontinuity of the same two-step process as (a)  in the
laboratory frame $q^+ > 0$.  The $u \bar u$ fluctuation of the
virtual photon first scatters inelastically on a neutron via a
single gluon exchange which produces an excited color state $8_C$
of the neutron.  This is then followed by the annihilation of the
$\bar u$ quark on a proton.  The two-step amplitudes of (b) or (c)
will interfere destructively with the single-step annihilation
amplitude on the proton alone, thus producing shadowing.  If the
proton spin $S^p_z$ is flipped ($\lambda^\prime \ne \lambda$) by
the valence interaction, as occurs in pseudoscalar Reggeon
exchange, then the single-step process cannot contribute to the
forward virtual Compton amplitude, and the two-step process itself
produces antishadowing of the valence quark distributions. Similar
processes occur in the case of the electroweak currents.

Figure~\ref{8686A02} illustrates a similar situation, but for the
three-gluon  ``Odderon" exchange.  In this case, the
two-step amplitudes of (b) or (c) can interfere constructively
with the Regge-behaved single-step annihilation amplitude on the
proton alone, thus producing antishadowing.   Similar processes
occur in the case of the electroweak currents.

\vspace{0.3cm}
\begin{figure}[htbp]
\begin{center}
\leavevmode {\epsfysize=10cm \epsffile{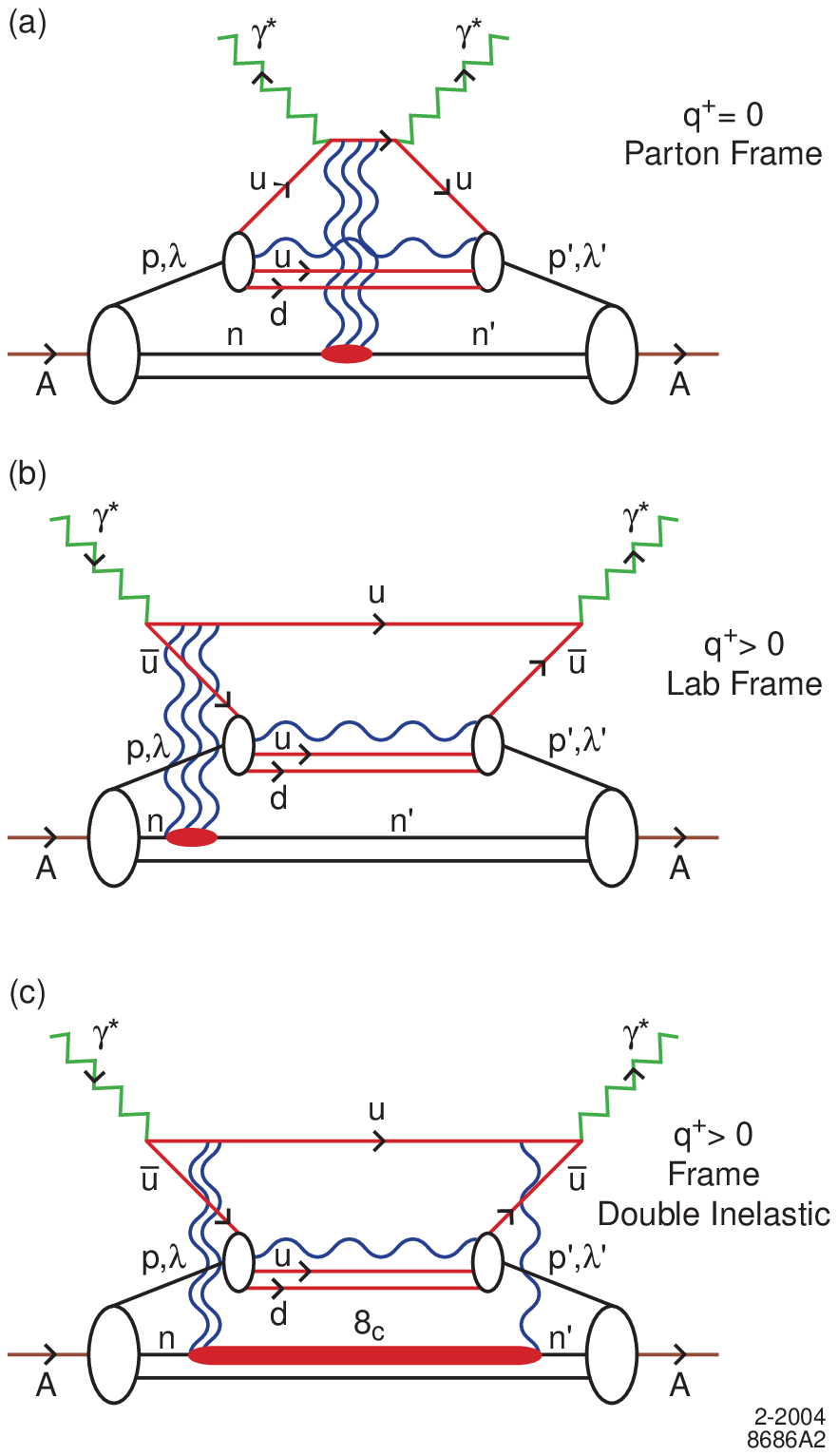}}
\end{center}
\caption[*]{\baselineskip 13pt Representation of leading-twist QCD
contributions to the nuclear structure function from the
absorptive part of the forward virtual Compton amplitude {\it
Im}$~ T(\gamma^* A \to \gamma^* A)$ involving odderon exchange.
(a) Illustration  of a two-step contribution in the $q^+ = 0,
q^2_\perp = Q^2$ parton model frame -- deep inelastic lepton
scattering on a valence quark of a target proton followed by the
final-state three-gluon exchange ``Odderon" interaction of the
outgoing quark on a target neutron. (b) Illustration of the
physics of the same two-step process shown in (a), but in the
laboratory frame where $q^+ > 0$.  The $u \bar u$ fluctuation of
the virtual photon first scatters elastically via three-gluon
exchange on a neutron; this is then followed by the annihilation
of the $\bar u$ quark on a proton. (c) Illustration of a doubly
inelastic discontinuity of the same two-step process as (a)  in
the laboratory frame $q^+ > 0$. The $u \bar u$ fluctuation of the
virtual photon first scatters inelastically on a neutron via
two-gluon exchange which produces an excited color state $8_C$ of
the neutron.  This is then followed by the annihilation of the
$\bar u$ quark on a proton. The two-step amplitudes of (b) or (c)
can interfere constructively with the Regge-behaved single-step
annihilation amplitude on the proton alone, thus producing
antishadowing.   Similar processes occur in the case of the
electroweak currents. \label{8686A02}}
\end{figure}

The Reggeon contributions to the quark scattering amplitudes
depend specifically on the quark flavor; for example the isovector
Regge trajectories couple differently to $u$ and $d$ quarks. The
$s$ and $\bar s$ couple to yet different Reggeons. This implies
distinct anti-shadowing effects for each quark and antiquark
component of the nuclear structure function; this in turn implies
nonuniversality of antishadowing of the charged, neutral, and
electromagnetic current. Anti-neutrino and neutrino reactions will
also have different antishadowing effects. In addition, there is
another source of antishadowing, specific to non-abelian theories,
which is discussed in more detail in the appendix.  It includes
one-gluon exchange~$\times$ Reggeon exchange, assuming the
existence of hidden-color components in the nuclear wavefunction.
We do not explicitly include this effect in our analysis since the
parameterization is uncertain.

There are also antishadowing contributions arising from two-step
processes involving Reggeon $\times$ Reggeon exchange, but these
contributions are power-law suppressed in the Bjorken limit.   We
will not include these higher-twist effects in our analysis.

In this paper we shall show in detail that the Gribov-Glauber
picture for nuclear deep inelastic scattering leads to
substantially different nuclear effects for charged and neutral
currents; in fact, the neutrino and antineutrino cross sections
are each modified in substantially different ways due to the
various allowed Regge exchanges.  This non-universality of nuclear
effects will modify the extraction of the weak-mixing angle
$\sin^2\theta_W$ obtained from the ratio of charged and neutral
current deep inelastic neutrino-nucleus scattering.

\section{Nuclear shadowing and antishadowing effects due to multiple scattering}

In this section, we will extend the analysis of
Ref.~\cite{BHPRL90} for the electromagnetic interaction case to
the neutrino DIS case.  The general approach is based on the
``covariant parton model~\cite{Landshoff:1970ff,Brodsky:1973hm},
which provides a relationship of  deep inelastic cross section to
quark-nucleon scattering. The central idea is the following: In
neutrino DIS on a nucleus A, although the virtual current may
interact inelastically with a nucleon coming from the nucleus in a
one-step process as shown in Fig.~\ref{bsy1f1}(a), it can also
interact elastically with several nucleons before the final
nucleon interacts inelastically as depicted in
Fig.~\ref{bsy1f1}(b). The interacting antiquark or quark is
spacelike:
\begin{equation}
\tau^2= -k^2 = x (s + k_{\bot}^2)/(1-x) -x M^2 +k_{\bot}^2
\end{equation}
is the negative of the invariant momentum squared of the interacting
parton.  Here
$M=\frac{1}{2}(M_p+M_n)$ is the nucleon mass, and $k_{\bot}$ is the
parton's transverse
momentum.  The quark-nucleon amplitude is assumed to be damped at large
quark virtuality
$\tau$, so that the quark-nucleon invariant $s = (k+p)^2$ grows as $1/x.$ This
description of deep inelastic scattering is consistent with recent analysis
of the role
of final state gluon interactions in QCD when one chooses light-cone gauge
to make the
Wilson line integral vanish~\cite{Brodsky:2002ue}.

At high energies the phase of the elastic amplitude is
approximately imaginary since it corresponds to Pomeron exchange.
The accumulated phase in multiple scattering is also imaginary.
Therefore, the two-step amplitude is coherent and opposite in
phase to the one-step amplitude where the beam interacts directly
on $N_2$ without initial-state interactions; the target nucleon
$N_2$ feels less incoming flux, which results in nuclear
shadowing.  Since there may be an $\alpha_R$ Reggeon or Odderon
contribution to the $\bar{q}N$ amplitude, the real phase
introduced by such contributions leads to antishadowing effect.
In this picture, antishadowing is attributed to a dynamical
mechanism rather than being enforced to satisfy the momentum sum
rule~\cite{Frankfurt}. The sum rule can still be maintained by the
nuclear modifications of the gluon distribution.

We now develop the detailed formulas which describe the nuclear
shadowing and antishadowing effects.

\subsection{Parameterizations of quark-nucleon scattering}

We shall assume that the high-energy antiquark-nucleon scattering
amplitude $T_{\bar{q}N}$ has the Regge and analytic behavior
characteristic of normal hadronic amplitudes.  Following the model
of Ref.~\cite{BHPRL90}, we consider a standard Reggeon at
$\alpha_R=\frac12$, an Odderon exchange term, a pseudoscalar
exchange term, and a term at $\alpha_R= -1$, in addition to the
Pomeron-exchange term.

The Pomeron exchange has the intercept $\alpha_P=1+\delta$.  For
the amputated $\bar{q}-N$  amplitude $T_{\bar{q}N}$ and $q-N$
amplitude $T_{qN}$  with $q=u$, and $d$, $N=p$, and $n$, we assume
the following parameterizations, including terms which represent
pseudoscalar Reggeon exchange.  Then resulting amplitudes are:
\begin{eqnarray}
T_{\bar{u}-p} &=& \sigma \Bigg[ s \left(i + \tan
\frac{\pi\delta}{2}\right) \beta_1 (\tau^2) - s \beta_{\cal O}
(\tau^2) - (1-i) s^{1/2} \beta_{1/2}^{0^+}(\tau^2)\\ \nonumber
&\quad& + (1+i) s^{1/2} \beta_{1/2}^{0^-}(\tau^2) - (1-i) s^{1/2}
\beta_{1/2}^{1^+}(\tau^2) + W (1-i) s^{1/2} \beta_{1/2}^{\rm
pseudo}(\tau^2) \\[1ex] &\quad&  + (1+i) s^{1/2}
\beta_{1/2}^{1^-}(\tau^2) + i s^{-1} \beta_{-1}^u(\tau^2)\Bigg],
\nonumber \label{Tu}
\end{eqnarray}
\begin{eqnarray}
T_{\bar{d}-p} &=& \sigma [ s \left(i + \tan \frac{\pi\delta}{2}\right)
\beta_1 (\tau^2)
-s \beta_{\cal O} (\tau^2) - (1-i) s^{1/2} \beta_{1/2}^{0^+}(\tau^2)\\
\nonumber
&\quad& + (1+i) s^{1/2} \beta_{1/2}^{0^-}(\tau^2)
 + (1-i) s^{1/2} \beta_{1/2}^{1^+}(\tau^2) + W (1-i) s^{1/2}
\beta_{1/2}^{\rm pseudo}(\tau^2)\\[1ex]
 &\quad&  - (1+i) s^{1/2}
\beta_{1/2}^{1^-}(\tau^2) + i s^{-1} \beta_{-1}^d(\tau^2)], \nonumber
\end{eqnarray}
\begin{eqnarray}
T_{u-p} &=& \sigma \Bigg[ s \left(i + \tan \frac{\pi\delta}{2}\right)
\beta_1 (\tau^2)
+ s \beta_{\cal O} (\tau^2) - (1-i) s^{1/2} \beta_{1/2}^{0^+}(\tau^2)\\
\nonumber
&\quad& - (1+i) s^{1/2} \beta_{1/2}^{0^-}(\tau^2)
 - (1-i) s^{1/2} \beta_{1/2}^{1^+}(\tau^2)
  + W (1-i) s^{1/2} \beta_{1/2}^{\rm pseudo}(\tau^2) \\[1ex]
  &\quad& - (1+i) s^{1/2}
\beta_{1/2}^{1^-}(\tau^2)\Bigg], \nonumber
\end{eqnarray}
\begin{eqnarray}
T_{d-p} &=& \sigma \Bigg[ s \left(i + \tan
\frac{\pi\delta}{2}\right)\beta_1 (\tau^2)
+s \beta_{\cal O} (\tau^2) - (1-i) s^{1/2} \beta_{1/2}^{0^+}(\tau^2)\\
\nonumber
&\quad& - (1+i) s^{1/2} \beta_{1/2}^{0^-}(\tau^2)
 + (1-i) s^{1/2} \beta_{1/2}^{1^+}(\tau^2)
 + W (1-i) s^{1/2} \beta_{1/2}^{\rm pseudo}(\tau^2) \\[1ex]
 &\quad& + (1+i) s^{1/2}
\beta_{1/2}^{1^-}(\tau^2)\Bigg],\nonumber
\end{eqnarray}
\begin{eqnarray}
T_{\bar{u}-n} &=& \sigma \Bigg[ s (i + \tan \frac{\pi\delta}{2}) \beta_1
(\tau^2)
-s \beta_{\cal O} (\tau^2) - (1-i) s^{1/2} \beta_{1/2}^{0^+}(\tau^2)\\
\nonumber
&\quad& + (1+i) s^{1/2} \beta_{1/2}^{0^-}(\tau^2)
 + (1-i) s^{1/2} \beta_{1/2}^{1^+}(\tau^2) + W (1-i) s^{1/2}
 \beta_{1/2}^{\rm pseudo}(\tau^2) \\[1ex]
 &\quad& - (1+i) s^{1/2}
\beta_{1/2}^{1^-}(\tau^2) + i s^{-1} \beta_{-1}^d(\tau^2)\Bigg],\nonumber
\end{eqnarray}
\begin{eqnarray}
T_{\bar{d}-n} &=& \sigma \Bigg[ s \left(i + \tan \frac{\pi\delta}{2}\right)
\beta_1 (\tau^2)
-s \beta_{\cal O} (\tau^2) - (1-i) s^{1/2} \beta_{1/2}^{0^+}(\tau^2)\\
\nonumber
&\quad& + (1+i) s^{1/2} \beta_{1/2}^{0^-}(\tau^2)
 - (1-i) s^{1/2} \beta_{1/2}^{1^+}(\tau^2)+
 W (1-i) s^{1/2} \beta_{1/2}^{\rm pseudo}(\tau^2) \\[1ex]
 &\quad& + (1+i) s^{1/2}
\beta_{1/2}^{1^-}(\tau^2) + i s^{-1} \beta_{-1}^u(\tau^2)\Bigg],\nonumber
\end{eqnarray}
\begin{eqnarray}
T_{u-n} &=& \sigma \Bigg[ s \left(i + \tan \frac{\pi\delta}{2}\right)
\beta_1 (\tau^2)
+ s \beta_{\cal O} (\tau^2) - (1-i) s^{1/2} \beta_{1/2}^{0^+}(\tau^2)\\
\nonumber
&\quad& - (1+i) s^{1/2} \beta_{1/2}^{0^-}(\tau^2)
 + + W (1-i) s^{1/2} \beta_{1/2}^{\rm pseudo}
 (\tau^2) \\[1ex]
 &\quad& + (1-i) s^{1/2} \beta_{1/2}^{1^+}(\tau^2) + (1+i) s^{1/2}
\beta_{1/2}^{1^-}(\tau^2)\Bigg],\nonumber
\end{eqnarray}
\begin{eqnarray}
T_{d-n} &=& \sigma \Bigg[ s \left(i + \tan \frac{\pi\delta}{2}\right)
\beta_1 (\tau^2)
+s \beta_{\cal O} (\tau^2) - (1-i) s^{1/2} \beta_{1/2}^{0^+}(\tau^2)\\
\nonumber
&\quad& - (1+i) s^{1/2} \beta_{1/2}^{0^-}(\tau^2)
 - (1-i) s^{1/2} \beta_{1/2}^{1^+}(\tau^2)
 + W (1-i) s^{1/2} \beta_{1/2}^{\rm pseudo}(\tau^2) \\[1ex]
 &\quad& - (1+i) s^{1/2}
\beta_{1/2}^{1^-}(\tau^2)\Bigg].\nonumber
\end{eqnarray}
$W=0$ and 1, since the pseudoscalar term cannot act just once in
the multiple scattering.

Here
\begin{equation}
\beta_j(\tau^2) = \frac {f_j} {1 + (\tau^2/ \bar{\nu}_j^2)^{n_j}} \label{beta}
\end{equation}
with $j=1$, $1/2$, $-1$, ${\cal O}$ and $pseudo$ (pseudoscalar). The parameters
$n_j$ ($j=1$, $1/2$ and $-1$) are taken to be the same as those in
Ref.~\cite{BHPRL90}. [See also Table I.]

The odd-C Odderon with $\alpha_{\cal O}=1$ has a real coupling
compared to the imaginary coupling of the even C Pomeron. It
reduces nuclear shadowing and produces antishadowing although it
does not contribute to the free nucleon structure functions. In
the  following numerical estimate, we take $f_{\cal O}=0.1$. In
order to fit the large $x$ experimental data on the parton
distributions of the nucleon, we introduce different values for
the parameters $\bar{\nu}_{-1}^{2}$ which control the off-shell
dependence of the ${\bar{q}-N}$ amplitudes. We denote them as
$\bar{\nu}_{-1}^{(u)2}$ and $\bar{\nu}_{-1}^{(d)2}$ for the $u$
and $d$ quarks, respectively.

We take the overall amplitude strength $\sigma$ to be the same in
all cases, with a value:
\begin{equation}
 \sigma=66 ~\rm{mb}.
\end{equation}
The $I=1$ Reggeon terms in the amplitudes play a very important
role, reflecting the sea asymmetry $\bar{d}-\bar{u}$ of the
nucleon in the low $x$ region.

In principle, $I=1$ pseudoscalar exchange should also contribute
here, but the $\gamma* A \to \gamma* A$ cross section is not
sensitive to its parameters, and therefore its strength cannot be
fixed. Careful fits to DVCS and other processes sensitive to the
$I=1$ pseudoscalar exchanges are needed. The $I=0$ coupling is
constrained by our fit to antishadowing for electromagnetic DIS.
Then we can predict antishadowing for weak DIS.

In principle, each Reggeon in the model of $q N$ scattering
amplitude  should couple to the individual quarks with the
appropriate isospin and charge conjugation dependence. For
example, the $\rho$ Reggeon couples as an $I=1$, $C=-$ exchange.
Although Reggeons of both $C = \pm$ appear in the quark-nucleon
amplitude,  in the end after multiple interactions and summing
over quark and antiquark currents, the nuclear Compton amplitude
has only $C=+$ exchange in the $t$-channel.  In our model, the
Reggeon term in the $q-N$ amplitude with $I=0$ and $C=+$ is taken
to represent the sum of the possible Reggeon exchanges. At leading
twist only the Pomeron and Odderon which derive from gluon
exchange survive in the multiple scattering. The Reggeon exchange
to elastic scattering in the multiple  scattering is suppressed.

In the present analysis, we also include the strange quark
contribution. The anti-strange quark can scatter elastically on
one nucleon via  Pomeron, Odderon and $\phi$ Reggeon exchanges.
The Reggeon intercept for the $\phi$ trajectory is close to
$\alpha_R(0) \sim 0$ since $\alpha_R(m^2_\phi)=1$ and the Regge
slope is universal. Actually, the $\phi$ trajectory can be
parameterized as $\alpha_R(t)=0.1+0.9 t $~\cite{Collins77}. Then
we can parameterize the amputated  $\bar{s}-N$ and $s-N$
amplitudes as:
\begin{equation}
T_{\bar{s}-N}= \sigma_{\bar{s}-N} [ i s \beta_1^{(s)} (\tau^2)-
s \beta_{\cal O}^{(s)} (\tau^2) + s^{0.1} ((1+\cos 0.1\pi)
-i \sin 0.1\pi)\beta_{0.1}(\tau^2) ],
\end{equation}
\begin{equation}
T_{s-N}= \sigma_{s-N} [ i s \beta_1^{(s)} (\tau^2)+
s \beta_{\cal O}^{(s)} (\tau^2) + s^{0.1} ((1+\cos 0.1\pi)
-i \sin 0.1\pi)\beta_{0.1}(\tau^2) ],
\end{equation}
where $\sigma_{\bar{s}-N}=\sigma_{s-N}=\sigma$ and $N=p$, and $n$.
Since the Pomeron coupling to the strange quark could be less in
strength than its coupling to light quarks, $f_1^{(s)}$ should be
smaller than $f_1=1$ for $u$ and $d$ quarks. We take
$f_1^{(s)}=0.1$. The value of the Reggeon coupling $f_{0.1}$ in
$\beta_{0.1}(\tau^2)$ is taken as 0.2 GeV$^{0.9}$, which is the
suitable mass dimension in order to have the proper mass dimension
of the Reggeon terms. With the above choice of the parameters, we
can produce a shape of strange quark distributions which is close
to those obtained by a fit analysis~\cite{Strange}, CTEQ-5
parametrization~\cite{CTEQ-5} and MRST
parametrization~\cite{MRST}.

If we use  $N=p$ and  $n$, indicating a proton and neutron target,
respectively, then for an isoscalar target $N_0$, the amplitude
$T_{\bar{q}N_0}$ per nucleon is
\begin{equation}
T_{\bar{q}N_0}=\frac{1}{2}
(T_{\bar{q}p} + T_{\bar{q}n} ).
\end{equation}
Then we introduce
\begin{equation}
T_{N_0}(s,\tau^2)=T_{\bar{q}N_0}
(s,\tau^2) \Delta_F^2(\tau^2),
\end{equation}
and
\begin{equation}
i\Delta_F (\tau^2) \sim \frac{1}{\bar{\nu}_p^2+\tau^2}.
\label{deltaf}
\end{equation}

Now let us turn to the scattering on a nuclear ($A$) target. We
expect that the $\bar{q}-A$ scattering amplitude can be obtained
from the $\bar{q}-N$ amplitude according to Glauber's theory as
follows,
\begin{equation}
T_{\bar{q}A}=\sum_{k_1=0}^{Z} \sum_{k_2=0}^{N}
\frac{1}{k_1+k_2} \left (
\begin{array}{c}
Z+N\\
k_1+k_2
\end{array}
\right )
\frac{1}{M} \alpha^{k_1+k_2-1}
\left (T_{\bar{q}p}\right )^{k_1}
\left (T_{\bar{q}n}\right )^{k_2}
 \theta (k_1 + k_2 -1)
\end{equation}
where
\begin{eqnarray}
M&=&\rm{Min}\{k_1+k_2, Z\}- \rm{Max}\{k_1+k_2-N, 0\} +1 \nonumber\\
 &=&\rm{Min}\{k_1+k_2, N\}- \rm{Max}\{k_1+k_2-Z, 0\} +1
\end{eqnarray}
and
\begin{equation}
\alpha=\frac{i}{4 \pi p_{c.m.} s^{1/2} (R^2 + 2 b)}
\end{equation}
with
\begin{equation}
p_{c.m.}=\sqrt{\tau^2 + (s -M^2-\tau^2)^2/4s},
\end{equation}
\begin{equation}
R^2=\frac23 R_0^2, R_0=1.123 A^{1/3} \rm{fm},
\end{equation}
and $b=10~(\rm{GeV/c})^{-2}$ is used~\cite{BHPRL90}. Furthermore,
we introduce
\begin{equation}
T_{A}(s,\tau^2)=T_{\bar{q}A}(s,\tau^2)
\Delta_F^2(\tau^2).
\end{equation}
Similar expressions hold for $T_{qA}$.

The Regge contribution to the deep inelastic cross section comes
from the handbag contribution to the forward virtual Compton
amplitude $\gamma^* p \to \gamma^* p.$ The Regge behavior
$x^{-\alpha_R(0)}$ arises from the summation over higher Fock
states.  The phase of the $I=0$ Reggeon contribution $(-i+1)$ with
$\alpha_R = 1/2$ entering the virtual Compton amplitude  is
opposite to the positive imaginary  contribution of pomeron
exchange and thus tends to reduce the deep inelastic cross section
on a nucleon.

As shown in Ref.~\cite{Brodsky:2002ue}, the multiple scattering
contributions from elastic scattering from Reggeon exchange is a
higher-twist contribution to the deep inelastic cross section;
only gauge interactions have a FSI effect in the Bjorken
limit~\cite{Brodsky:2002ue}. The Pomeron and Odderon nominally
have $\alpha_{\cal O} \simeq 1,$  so their contributions to
elastic scattering are not suppressed in the Bjorken limit, since
they are derived from multiple gluon exchange. Thus the Pomeron
and Odderon can act any number of times in the nucleus, but the
Reggeon can act only once at leading twist. In effect the Reggeon
does not have enough time to form in the FSI at small invariant
separation $x^2 \sim 1/Q^2$. Thus FSIs from Reggeons in $T_{\bar q
A}$ with $\alpha_R \sim 0.5 $ should be suppressed in the Bjorken
limit by a power of $1/Q.$ In order to implement this we put a
suppression factor $R_D$ in the multiple scattering Reggeon terms:
\begin{equation}
R_D= \left(\frac{Q_0^2}{Q_0^2 + Q^2} \right )^{1/2}
\end{equation}
for the $u$ and $d$ quarks. And
\begin{equation}
R_D= \left(\frac{Q_0^2}{Q_0^2 + Q^2} \right )
\end{equation}
for the $s$ quark with $Q_0^2\simeq 1~\rm{GeV}^2$, a typical hadronic scale.

When we take the limit of large $Q^2$, the antishadowing due to
elastic Reggeon exchanges is quenched; however, the presence of
the Odderon can produce anti-shadowing. For example, a two-step
nuclear process shown in Fig.~3(b) from elastic Odderon scattering
plus inelastic Reggeon scattering gives a contribution to the
virtual Compton amplitude $1 \times i \times (i + 1) = (-1 + i).$
[The middle factor of $i$ is due to the cut between the two
steps.] The positive imaginary contribution to the two step
amplitude produces an enhancement of the nuclear cross section
relative to the nucleon cross section in the regime $x \sim 0.1 $
where the Reggeon contribution to deep inelastic scattering is
important. This is a key feature of our model. Note that the
two-step process of Odderon plus Pomeron produces only a real
contribution to the virtual Compton amplitude. The Pomeron-Pomeron
and Pomeron-Reggeon two-step contributions reduce the one-step
Pomeron plus Reggeon contributions, respectively, and thus only
produce nuclear shadowing.

We can also consider the two-step  [$ P i R_\pi$] contribution to
${\it Im}~ T_{\gamma^* A \to \gamma^* A}$, which involves the
imaginary part ${\it Im}~ R_\pi$ of the (nonforward) pion Reggeon
exchange amplitude. The pion pole term alone is real so we
consider the pion Regge trajectory---the Reggeized version of pion
exchange.  When we take its absorptive part, we look at the cut
through a $q \bar q$ ladder exchanged on the second nucleon. Since
the Reggeized pion exchange has $I=1$ and the pomeron is $I=0$,
the two-step  [$ P i R_\pi$]  contribution does not contribute to
${\it Im}~ T_{\gamma^* A \to \gamma^* A}$ if the target is $ I=0$
Thus there is no antishadowing contribution from [$ P i R_\pi$]
to the deuteron structure function. However, there are other
pseudoscalar exchanges possible, such as the $\eta$ Reggeon.

It is interesting to analyze the situation form the point of view
of angular momentum. The question is whether the two step process
[$ P i R_\pi$] require orbital angular momentum in the ground
state nuclear wavefunction. Consider the  [$ P i R_\pi$]
contribution to the forward virtual Compton amplitude ${\it Im}~
T_{\gamma^* A \to \gamma^* A}.$ The pomeron exchange on the first
nucleon gives a transverse momentum kick $\vec k_\perp$. The pion
Reggeon exchange on the second nucleon gives a balancing opposite
kick $-\vec k_\perp$ so that we can have a forward nuclear
amplitude. The pseudoscalar exchange on the second nucleon is a
$\Delta L=1$ transition. That is why the amplitude requires
nonzero $k_\perp.$ Thus we are actually looking at the overlap of
nuclear Fock components of the nucleus with $\Delta L =1.$  This
is the same admixture which in the spin-1/2 case which gives a
nuclear magnetic anomalous moment.

The unpolarized quark distribution functions  in an isoscalar
target ($N_0$) and nucleus target ($A$) are, respectively,
\begin{eqnarray}
xq^{N_0}(x)&=&\frac {2} {(2\pi)^3} \frac{C x^2}{1-x}
\int ds d^2 {\bf{k}}_{\bot}
\rm{Im} T_{N_0}(s,\mu^2),\\[1ex]
xq^{A}(x)&=& \frac {2} {(2\pi)^3} \frac{C x^2}{1-x}
\int ds d^2 {\bf{k}}_{\bot}
\rm{Im} T_{A}(s,\mu^2),\\[1ex]
\mu^2&=& -\tau^2.
\end{eqnarray}
The constant $C$ is related to the parton wave
function renormalization constant.

With the obtained quark distributions $xq^{N_0(A)}(x)$ for
an isoscalar target $N_0$ or a nucleus target $A$, we can calculate
the structure functions for various current exchanges.

(1) The photon exchange case
\begin{eqnarray}
F_1^{\gamma N_0(A)} &=& \frac12
\Bigg\{
\frac{4}{9}
\left[u(x)^{N_0(A)} +\bar{u}^{N_0(A)}
\right]
+\frac{1}{9}
\bigg[d(x)^{N_0(A)} +\bar{d}^{N_0(A)}\\[1ex] \nonumber
&\quad& + s(x)^{N_0(A)} +\bar{s}^{N_0(A)}
\bigg]
\Bigg\},\\[1.5ex]
F_2^{\gamma N_0(A)}  &=& 2 x F_1^{\gamma N_0(A)}.
\end{eqnarray}

(2) The neutral current exchange case

The structure functions of the NC reaction  are
\begin{eqnarray}
F_1^{Z N_0(A)} &=& \frac12
\Bigg\{ \left[(g_V^u)^2 + (g_A^u)^2\right] (u^{N_0(A)}(x)
+ \bar{u}^{N_0(A)}(x))\\ \nonumber
&\quad& +  [(g_V^d)^2 + (g_A^d)^2] \left(d^{N_0(A)}(x) + \bar{d}^{N_0(A)}(x) +
s^{N_0(A)}(x) +\bar{s}^{N_0(A)}(x)\right)\Bigg\},  \\
F_2^{Z N_0(A)}  &=& 2 x F_1^{Z N_0(A)},
\label{FiNC}\\[1.5ex] \nonumber
F_3^{Z N_0(A)} &=& 2 \Bigg[ g_V^u g_A^u ( u^{N_0(A)}(x) - \bar{u}^{N_0(A)}(x) )
\\[1ex]
 &\quad& + g_V^d g_A^d (d^{N_0(A)}(x) - \bar{d}^{N_0(A)}(x) +
s^{N_0(A)}(x) -\bar{s}^{N_0(A)}(x))\Bigg].
\end{eqnarray}
In the SM the vector and axial-vector quark couplings are given by
\[ g_V^u=\frac12-\frac43 \sin^2\theta_W, \ \ \
g_V^d=-\frac12+\frac23 \sin^2\theta_W, \ \ \ \ g_A^u=\frac12,\ \ \
\  g_A^d=-\frac{1}{2}, \]
where $\sin^2\theta_W$ is the weak-mixing angle.

(3) The charged current exchange case

The structure function of the CC reaction is given by
\begin{eqnarray} \label{F1WP}
F_1^{W^+ N_0(A)} &=& \bar{u}^{N_0(A)}(x) (|V_{ud}|^2 + |V_{us}|^2) +
\bar{u}^{N_0(A)}(\xi_b)|V_{ub}|^2 \theta(x_b-x)\\[1ex] \nonumber
&\quad& + d^{N_0(A)}(x)|V_{ud}|^2 + d^{N_0(A)}(\xi_c)|V_{cd}|^2
\theta(x_c-x)\\[1ex]
\nonumber
&\quad& + s^{N_0(A)}(x)|V_{us}|^2 + s^{N_0(A)}(\xi_c)|V_{cs}|^2\theta(x_c-x),
\end{eqnarray}
here $V_{ij}$ are the Cabibbo-Kobayashi-Maskawa mixing matrix
elements. The variable
$$
   \xi_k =\left \{
      \begin{array}{ll}
      x \left( 1 +\frac{m_k^2}{Q^2} \right), & (k=c, b),\\
      x, & (k=u, d, s),
      \end{array} \right.
$$
and the step functions $\theta(x_c-x), \theta(x_b-x)$ take into
account rescaling due to heavy quark production thresholds.

The structure functions $F_2^{W^+ N_0(A)}$ and $F_3^{W^+ N_0(A)}$
are obtained from  (\ref{F1WP}) by the replacement of the quark
distribution functions $q(x, Q^2)$ indicated in the curly
brackets:
\begin{eqnarray}
F_2^{W^+ N_0(A)}(x, Q^2)
&=& F_1^{W^+ N_0(A)}(x, Q^2)
 \{q^{N_0(A)}(x, Q^2) \to 2x q^{N_0(A)}(x, Q^2), \\ \nonumber
&\quad &  \ q^{N_0(A)}(\xi_k, Q^2) \to  2 \xi_k q^{N_0(A)}(\xi_k,
Q^2)\},\label{F3} \\[1.5ex] F_3^{W^+ N_0(A)}(x, Q^2) &=& 2 \
F_1^{W^+ N_0(A)}(x, Q^2)\{\bar{q}^{N_0(A)}(x, Q^2) \to
-\bar{q}^{N_0(A)}(x, Q^2)\}.
\end{eqnarray}

There are similar formulas for the $W^-$-current exchange reaction.

\subsection{The values of the parameters}

In the last section we presented the formulas involved in our
formalism. We shall take the value for most of the parameters to
be the same as those in Ref.~\cite{BHPRL90}. The values of other
parameters are chosen in order to fit the experimental
data~\cite{NMC96,NMC91,F2-data} on $F_2^{N_0}$,$(F_2^p-F_2^n)$,
$F_2^n/F_2^p$; they are then checked against the known nuclear
shadowing and antishadowing effects~\cite{Sha1,Sha2}. A summary of
the set of parameters is given in Table~\ref{Table1}.

\begin{center}
\begin{table*}[h]
\caption{Parameters in our numerical calculation} \vspace{.6cm}
\begin{tabular}{|c|c||c|c|}\hline
$\bar{\nu}_{1}^2$ & 0.2 GeV$^2$ &$f_{1}$, $f_{1}^s$  & 1.0, 0.5 \\ \hline
$\bar{\nu}_{1/2}^2$ & 0.2  GeV$^2$&  $f_{1/2}^{0^+}$,
$f_{1/2}^{0^-}$, $f_{1/2}^{1^+}$, $f_{1/2}^{1^-}$ & 0.30,0.3,0.1,0.3 GeV
\\ \hline
$\bar{\nu}_{-1}^{(u)2}$,~~$\bar{\nu}_{-1}^{(d)2}$ & 1.3,~~0.65  GeV$^2$&
  $f_{-1}$ & 0.45 GeV$^4$\\ \hline
$\bar{\nu}_{\cal O}^2$ & 0.30 GeV$^2$ & $f_{\cal O}$ & 0.10  \\
\hline $\bar{\nu}_{p}^2$ & 1.0 GeV$^2$  & b & 10 (GeV/c)$^{-2}$ \\
\hline $~~~~~~~\sigma~~~~~~~$ & ~~~66~mb~~~~  & $n_{-1}$ & 2 \\
\hline $f_{\rm pseudo}$ & 1.35  GeV &  $n_1$, $n_{1/2}$, $n_0$ & 4\\
\hline
\end{tabular}\label{Table1}
\end{table*}
\end{center}

With the above parameters, the average nucleon structure function
\begin{equation}
F_2 = F_2^{N_0}=\frac{F_2^p +F_2^n}{2}
\end{equation}
for the photon exchange case are shown as a solid curve in
Fig.~\ref{bsy1f2}. The valence and sea contributions to $F_2$ are
also presented as dashed and dotted curves in Fig.~\ref{bsy1f2}.
$F_2$ is close to the SLAC and NMC~\cite{F2-data} experimental
data. We also show our results of the difference $F_2^p-F_2^n$ and
the ratio $F_2^n/F_2^p$ of the nucleon structure functions in
Fig.~\ref{bsy1f3} and Fig.~\ref{bsy1f4}, respectively. In the
coming subsection we will show the nuclear effects on the
structure functions and in the next section estimate the nuclear
shadowing and antishadowing effects on the extraction of
$\sin^2\theta_W$.

\vspace{0.3cm}
\begin{figure}[htb]
\begin{center}
\leavevmode {\epsfysize=8cm \epsffile{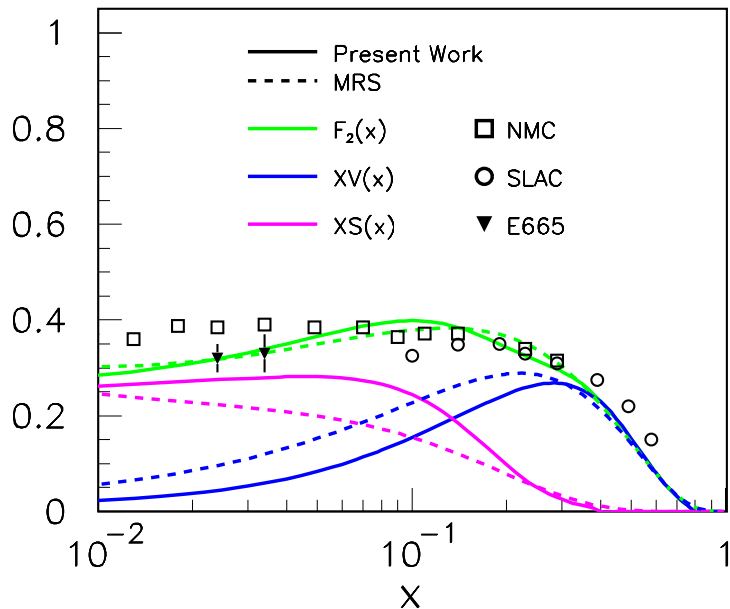}}
\end{center}
\caption[*]{\baselineskip 13pt The calculated nucleon structure
function $F_2$ (biggest solid curve), and valence  (next solid
curve at large $x$ values) and sea (smallest solid curve at large
$x$ values) contributions to $F_2$, and the corresponding result
of the MRST parametrization~\cite{MRST} (dashed curves), at $Q^2 =
1~\rm{GeV}^2$. The experimental data are taken from
Ref.~\cite{F2-data}.
 \label{bsy1f2}}
\end{figure}

\begin{figure}[htbp]
\begin{center}
\leavevmode {\epsfysize=8cm \epsffile{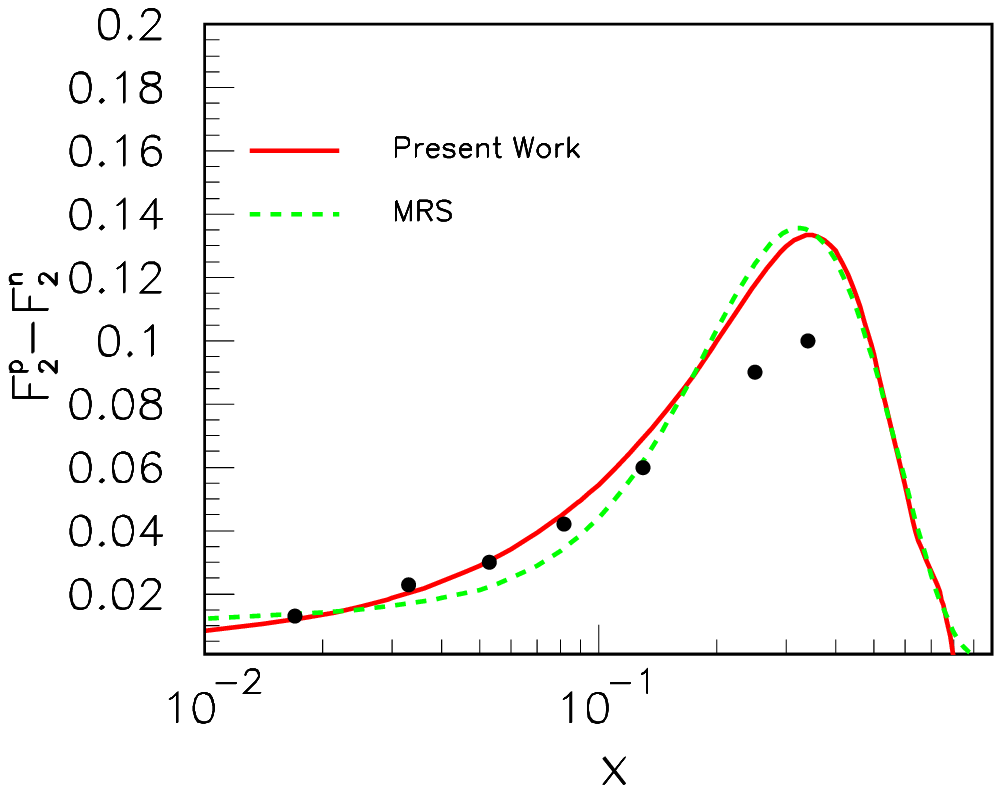}}
\end{center}
\caption[*]{\baselineskip 13pt The calculated difference
$F_2^p-F_2^n$ of the nucleon structure functions (solid curve),
and the corresponding result of the MRST
parametrization~\cite{MRST}(dashed curve), at $Q^2 =
1~\rm{GeV}^2$. The experimental data are taken from
Ref.~\cite{NMC91}.
 \label{bsy1f3}}

\vspace{0.3cm}
\begin{center}
\leavevmode {\epsfysize=6.5cm \epsffile{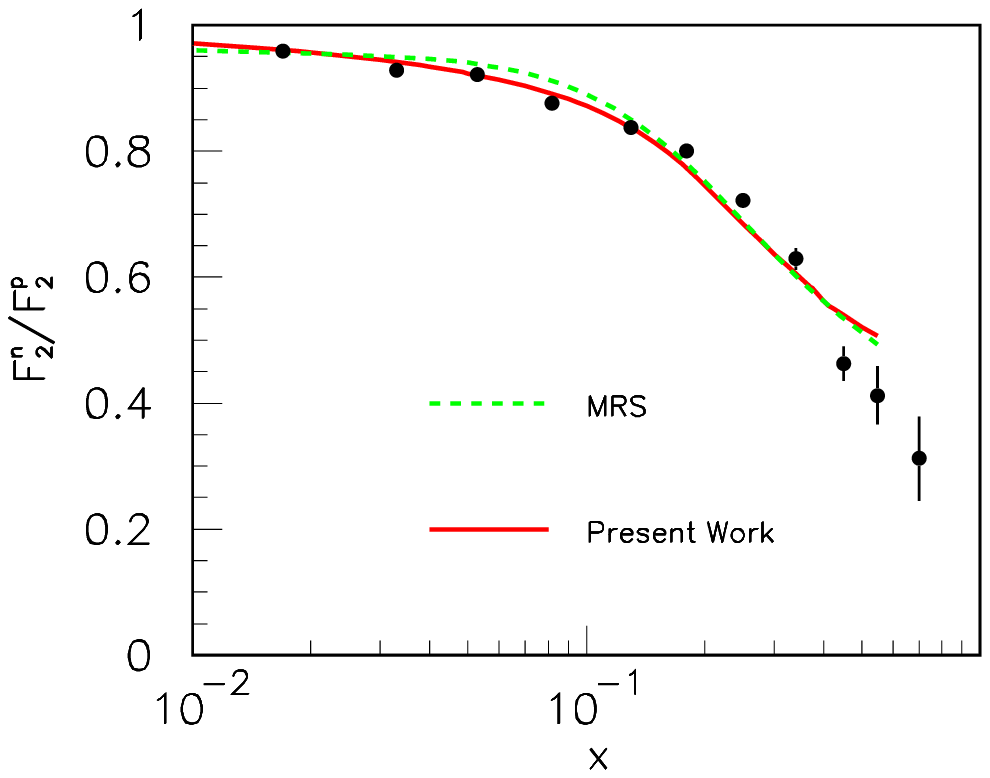}} 
\end{center}
\caption[*]{\baselineskip 13pt
 The calculated ratio  $F_2^n/F_2^p$
of the nucleon structure
functions (solid curve), and the corresponding result of the MRST
parametrization~\cite{MRST} at energy scale 1 GeV (dashed curve).
The experimental data are taken from Ref.~\cite{NMC91}.
\label{bsy1f4}}
\end{figure}

\vspace{0.3cm}
\begin{figure}[htb]
\begin{center}
\leavevmode {\epsfysize=10cm \epsffile{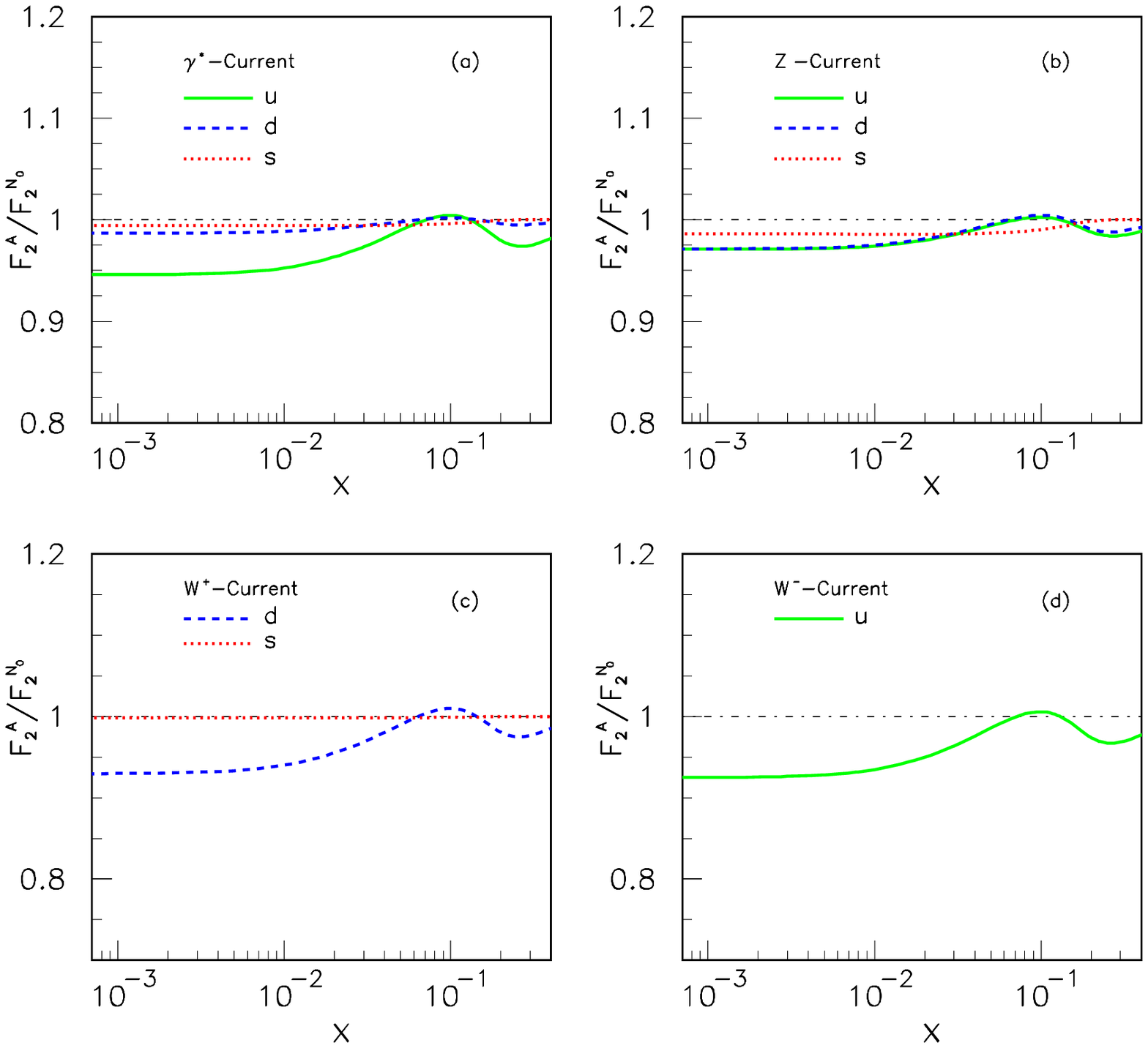}}
\end{center}
\caption[*]{\baselineskip 13pt
 The  quark contributions to the
ratios of structure functions at $ Q^2 = 1~\rm{GeV}^2$. The solid,
dashed and dotted curves correspond to the $u$, $d$ and $s$ quark
contributions, respectively. This corresponds in our model to the
nuclear dependence of the $\sigma(\bar u-A)$, $\sigma(\bar d-A)$,
$\sigma(\bar s-A)$ cross sections, respectively. In order to
stress the individual contribution of quarks, the numerator of the
ratio $F_2^{A} / F_2^{N_0}$ shown in these two figures is obtained
from the denominator by a replacement $q^{N_0}$ into $q^{A}$ for
only the considered quark. As a result, the effect of
antishadowing appears diminished.
 \label{bsy1f5}}
\end{figure}

\subsection{Nuclear shadowing and antishadowing effects}

We introduce the ratio
\begin{eqnarray}
R=F_2^{A} / F_2^{N_0}
\end{eqnarray}
to indicate the nuclear electromagnetic shadowing and
antishadowing effect. We will focus on the nucleus $^{56} Fe$
since an iron target was used in the NuTeV experiment and test the
nuclear effect in the $x>0.01$ region since $97\%$ of the NuTeV
data is from $0.01 < x < 0.75$~\cite{NuTeVx}. In
Figs.~\ref{bsy1f5}--\ref{bsy1f6}, we show the quark  $q$ and
anti-quark $\bar{q}$ contributions to the ratio of the structure
functions. In order to stress the individual contribution of
quarks, the numerator of the ratio $F_2^{A} / F_2^{N_0}$ shown in
these two figures is obtained from the denominator by a
replacement $q^{N_0}(\bar{q}^{N_0})$ into $q^{A}(\bar{q}^{A})$ for
only the considered quark (anti-quark). Because the strange quark
distribution is much smaller than $u$ and $d$ quark distributions,
the strange quark contribution to the ratio is very close to 1
although $s^{A}/s^{N_0}$ may significantly deviate from 1.

\vspace{0.3cm}
\begin{figure}[ht]
\begin{center}
\leavevmode {\epsfysize=11cm \epsffile{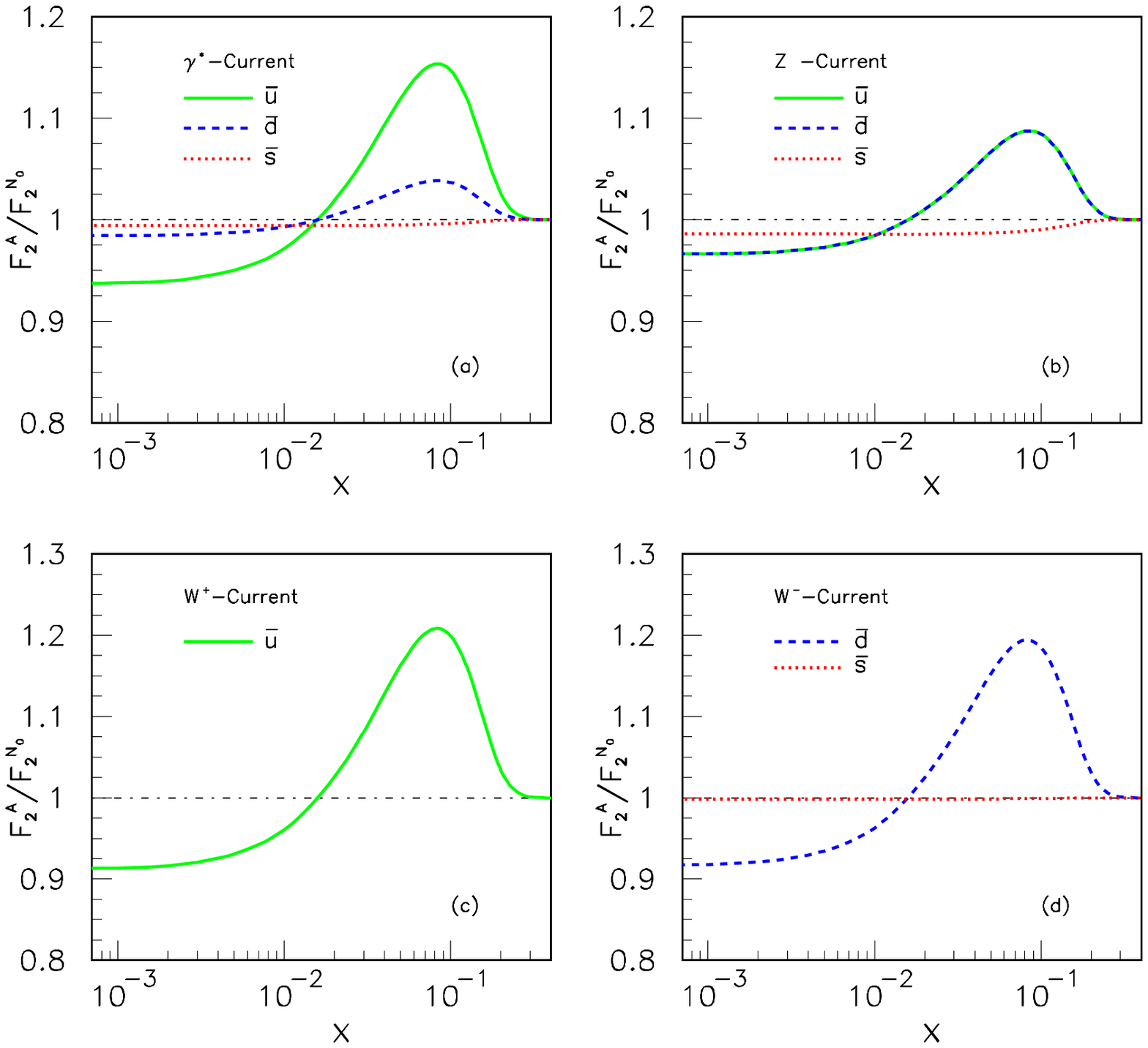}} 
\end{center}
\caption[*]{\baselineskip 13pt
 The anti-quark contributions to
ratios of the structure functions at $ Q^2 = 1~\rm{GeV}^2$. The
solid, dashed and dotted curves correspond to $\bar{u}$, $\bar{d}$
and $\bar{s}$ quark contributions, respectively. This corresponds
in our model to the nuclear dependence of the $\sigma(u-A)$,
$\sigma(d-A)$, $\sigma(s-A)$ cross sections, respectively. In
order to stress the individual contribution of quarks, the
numerator of the ratio $F_2^{A} / F_2^{N_0}$ shown in these two
figures is obtained from the denominator by a replacement
$\bar{q}^{N_0}$ into $\bar{q}^{A}$ for only the considered
anti-quark.
 \label{bsy1f6}}
\end{figure}

In Fig.~\ref{bsy1f7}, we give our prediction of the nuclear
shadowing and antishadowing effects (the sum of all quark and
antiquark contributions to the ratio $R$) for nuclei $^{56} Fe$
and $^{40} Ca$. From Fig.~\ref{bsy1f7}, we find that our model can
explain well the experimental data on the nuclear shadowing and
antishadowing effect in DIS for electromagnetic currents.

\vspace{0.3cm}
\begin{figure}[ht]
\begin{center}
\leavevmode {\epsfysize=8cm \epsffile{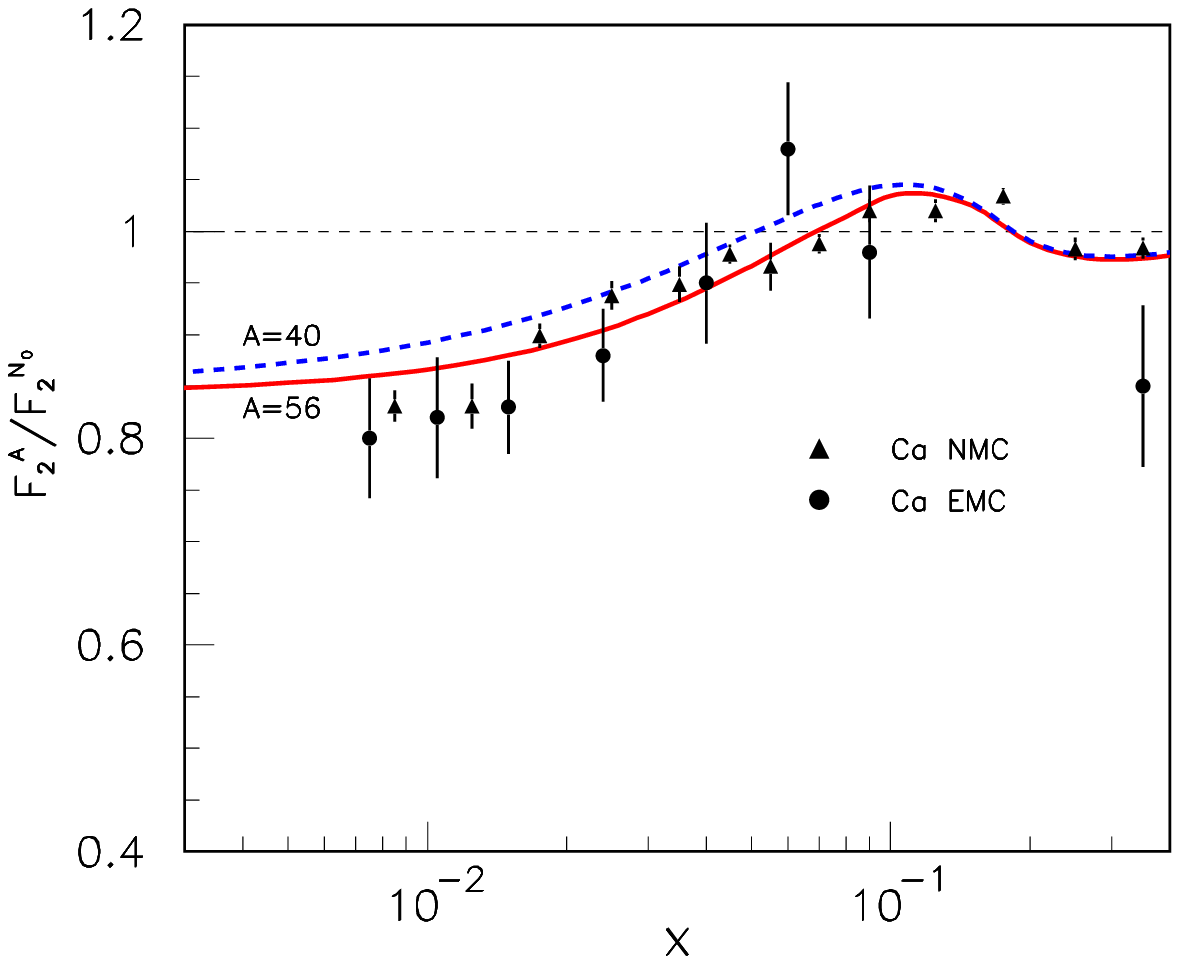}}
\end{center}
\caption[*]{\baselineskip 13pt
 The nuclear shadowing and antishadowing effects at $\langle
Q^2\rangle=1~\rm{GeV}^2$. The experimental data are taken from
Refs.~\cite{Sha1,Sha2}.
 \label{bsy1f7}}
\end{figure}

We can further check the our model by predicting the ratio of
$F_2$ structure functions
$F_{2A}\rm{[neutrino]}/(18/5)F_{2A}\rm{[muon]}$, which has been
measured by the NuTev collaboration~\cite{Yang:2000ju}. The
results are shown in Fig.\ref{8705A01}, and they agree very well
with the experimental data, in a calculation with no further free
parameters. Notice that the data for the ratio tends to go below
$1$ for $x > 0.4,$ which is also what we find. In this case we
have taken $Q^2 = 20~\rm{GeV}^2$, which is the average value of
the NuTeV experiment.

\vspace{0.3cm}
\begin{figure}[ht]
\begin{center}
\leavevmode {\epsfysize=8cm \epsffile{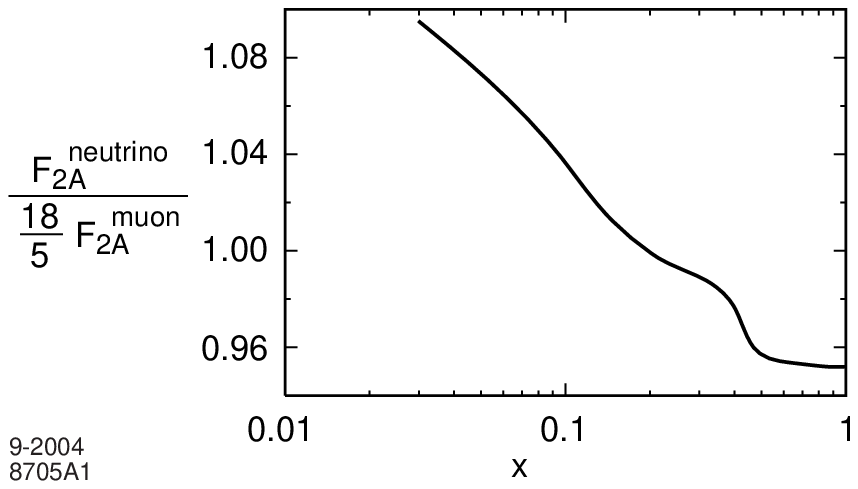}}
\end{center}
\caption[*]{\baselineskip 13pt Our prediction for the ratio of
$F_2$ structure functions
$F_{2A}\rm{[neutrino]}/(18/5)F_{2A}\rm{[muon]}$, measured in Ref.
~\cite{Yang:2000ju}, at $Q^2 = 20~\rm{GeV}^2$.
 \label{8705A01}}
\end{figure}

We emphasize that  the nuclear shadowing and antishadowing of the
different currents are not universal since they depend on the
different quark species. We still have factorization in the sense
that we will have the same shadowing quark by quark in nuclear
Drell-Yan processes.

In the case of weak currents, the shadowing/antishadowing effects
are strongly influenced by the behavior of the structure function
$F_3$, which is not present in the electromagnetic case. We will
present in the next section cross section ratios (nucleus/nucleon)
to illustrate the shadowing/antishadowing effects in weak current
interactions.

\section{Nuclear effects on extraction of $\sin^2\theta_W$}

The observables measured in neutrino DIS experiments are the
ratios of neutral current (NC) to charged current (CC) current
events; these are related via Monte Carlo simulations to
$\sin^2\theta_W$. In order to examine the possible impact of
nuclear shadowing and antishadowing corrections on the extraction
of $\sin^2\theta_W$, one is usually interested in the following
ratios
\begin{eqnarray}
R^\nu_A &=& \frac{\sigma (\nu_\mu + A \to \nu_\mu + X)}
{\sigma (\nu_\mu + A \to \mu^- + X)} \label{Rnu},\\[1.5ex]
R^{\bar{\nu}}_A &=& \frac{\sigma (\bar{\nu}_\mu + A \to \bar{\nu}_\mu + X)}
{\sigma (\bar{\nu}_\mu + A \to \mu^+ + X)} \label{Rnubar}
\end{eqnarray}
of NC to CC neutrino (anti-neutrino) cross sections for a nuclear
target A. As is well known, if nuclear effects are neglected for
an isoscalar target, one can extract the weak-mixing angle by
using the Llewellyn-Smith relation~\cite{LS}:
\begin{eqnarray}
R_N^{\ \nu[\bar{\nu}]} = \frac{\sigma (\nu_\mu[\bar\nu_{\mu}] + N
\to \nu_\mu[\bar\nu_{\mu}] + X)} {\sigma (\nu_\mu[\bar\nu_{\mu}] +
N \to \mu^-[\mu^+] + X)} = \rho_0^2 \left(\frac{1}{2}
-\sin^2\theta_W + \frac{5}{9} \sin^4 \theta_W ( 1+
r^{[-1]})\right),
\end{eqnarray}
written in terms of NC and CC (anti-)neutrino-nucleon cross
sections. Here,
\begin{equation}
\label{rho}
\rho_0 = \frac{M_W^2}{\cos^2\theta_W M_Z^2}, \ \ \ \ \
r = \frac{\sigma (\bar\nu_{\mu} + N \to \mu^+ + X)}
{\sigma (\nu_\mu + N \to \mu^- + X)}  \sim \frac{1}{2}.
\end{equation}
However, actual targets such as the iron target of the NuTeV
experiment are not always isoscalar, having a significant neutron
excess. In addition, as we have stressed here, nuclear effects due
to multi-scattering could be very important. These nuclear effects
should also modify the CC and NC structure functions, and
therefore a detailed study of these effects on the extraction of
the weak-mixing angle is essential. In order to reduce the
uncertainties related to sea quarks, Paschos and Wolfenstein
~\cite{PW} showed that one can extract $\sin^2\theta_W$ from the
relationship
\begin{eqnarray}
R_N^{^-} = \frac{\sigma (\nu_\mu + N \to \nu_\mu + X) - \sigma
(\bar\nu_{\mu} + N \to \bar\nu_{\mu} + X)} {\sigma (\nu_{\mu} + N
\to \mu^- + X) - \sigma (\bar\nu_{\mu} + N \to \mu^+ + X)} =
\rho_0^2 \left(\frac{1}{2} - \sin^2\theta_W\right).
\end{eqnarray}
Inspired by the above relation, we will  examine nuclear effects
on $\sin^2\theta_W$ by the following observable for the scattering
off a nuclear target $A$,
\begin{eqnarray}
R_A^{^-} = \frac{\sigma (\nu_\mu + A \to \nu_\mu + X) -
\sigma (\bar\nu_{\mu} + A \to \bar\nu_{\mu} + X)}
{\sigma (\nu_{\mu} + A \to \mu^- + X) -
\sigma (\bar\nu_{\mu} + A \to \mu^+ + X)}.
\end{eqnarray}

\vspace{0.3cm}
\begin{figure}[ht]
\begin{center}
\leavevmode {\epsfysize=10cm \epsffile{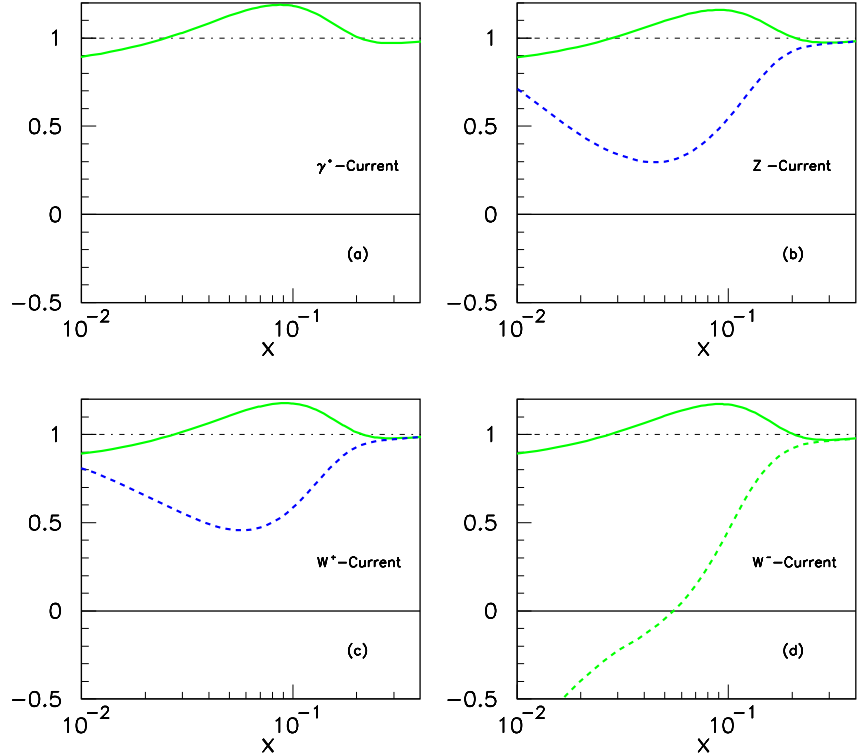}}
\end{center}
\caption[*]{\baselineskip 13pt
 Ratios $F_2^{A}/F_2^{N^0}$ (solid curves) and
$F_3^{A}/F_3^{N^0}$ (dashed curves)  for various current exchange
interactions, at $Q^2 = 1~\rm{GeV}^2$. \label{bsy1f8}}
\end{figure}

In the previous section, we have shown in Fig.~\ref{bsy1f7} the
nuclear effect on the electromagnetic structure functions. Here we
can also look the nuclear effect on the cross sections in CC and
NC neutrino-nucleus DIS. In Fig.~\ref{bsy1f8}, we show ratios
$F_2^{A}/F_2^{N^0}$ (solid curves) and $F_3^{A}/F_3^{N^0}$ (dashed
curves) for various current exchange interactions. The fact that
the $F_3^{A}/F_3^{N^0}$ ratio for the $W^-$-current becomes
negative and divergent for small $x$ comes from the behavior of
$F_3^{N^0}$, which in our model vanishes for $x \sim 0.01$. In
addition, we are interested in the following ratios
\begin{equation}
R^{\nu}_{Z}(x)= \frac{d\sigma (\nu_\mu + A \to \nu_\mu + X)/dx}
{d\sigma (\nu_\mu + N \to \nu_\mu + X)/dx},\label{RZ11}
\end{equation}
\begin{equation}
R^{\bar{\nu}}_{Z}(x)= \frac{d\sigma (\bar{\nu}_\mu + A \to \bar{\nu}_\mu +
X)/dx}
{d\sigma (\bar{\nu}_\mu + N \to \bar{\nu}_\mu + X)/dx},\label{RZ12}
\end{equation}
\begin{equation}
R^{\nu}_{W^+}(x)= \frac{d\sigma (\nu_{\mu} + A \to \mu^- + X)/dx}
{d\sigma (\nu_{\mu} + N \to \mu^- + X)/dx},\label{RW21}
\end{equation}
\begin{equation}
R^{\bar{\nu}}_{W^-}(x)= \frac{d\sigma (\bar\nu_{\mu} + A \to \mu^+ + X)/dx}
{d\sigma (\bar\nu_{\mu} + N \to \mu^+ + X)/dx}.\label{RW22}
\end{equation}
The above ratios are closely related to the nuclear effects in the
ratio,
\begin{equation}
R_{A/N}^-=R_A^{^-}/R_N^{^-}
\end{equation}
which are later used to extract the nuclear effect on the
weak-mixing angle. In Fig.~\ref{bsy1f9}, we show the ratios of
Eqs.~(\ref{RZ11})--(\ref{RW22}). From
Figs.~\ref{bsy1f7}--\ref{bsy1f9}, one finds that the nuclear
effect for charged and neutral currents is substantially different
from that for the electromagnetic nuclear structure functions.
There is a strong antishadowing effect in $R^{\bar{\nu}}_{Z}$ and
$R^{\bar{\nu}}_{W^-}$, but there is a small one in $R^{\nu}_{Z}$
and $R^{\nu}_{W^+}$. Moreover, for neutrinos the NC and CC
shadowing/antishadowing effects are the same, but for
antineutrinos they are substantially different.  As a result, in
the neutrino case the Llewellyn-Smith relation can be used in
order to extract the weak mixing angle, but it cannot be used for
this purpose in the case of antineutrino deep inelastic scattering
in nuclei.

If the nuclear target had zero isospin, and there was no
contribution from $s$ or $\bar s$ quarks, then there would be no
nuclear correction to the Llewellyn Smith relation. However, when
one includes the strange quark currents, the situation is
different. The neutral current interactions of the antineutrino on
the $\bar s$ are shadowed rather than antishadowed (see Fig.
\ref{bsy1f8}) in the region $x \sim 0.1,$ which reduces the total
antishadowing effect. The charged current interactions of the
antineutrino still experience strong antishadowing from the $\bar
d$.  In contrast, neutrino interactions are relatively insensitive
to the strange quark current since they are dominated by
interactions on the valence quarks, not the anti-quarks. Even when
one includes the strange quarks, antineutrino NC and CC
interactions experience more antishadowing than neutrinos (see
Fig. \ref{bsy1f9}).

\vspace{0.3cm}
\begin{figure}
\begin{center}
\leavevmode {\epsfysize=8cm \epsffile{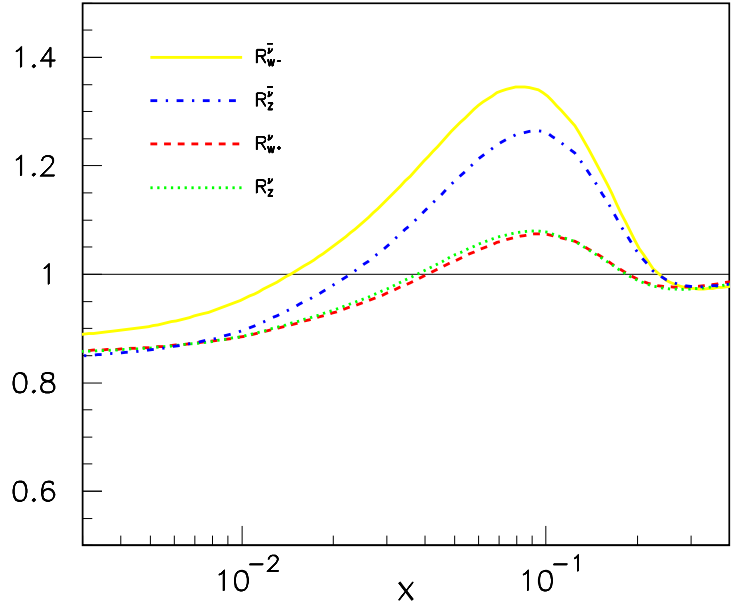}}
\end{center}
\caption[*]{\baselineskip 13pt The  nuclear effect on the cross
sections of CC and NC neutrino-nucleus DIS, at $Q^2 =
1~\rm{GeV}^2$. The dotted  and dashed curves almost overlap.
 \label{bsy1f9}}
\end{figure}

\begin{figure}
\begin{center}
\leavevmode {\epsfysize=8cm \epsffile{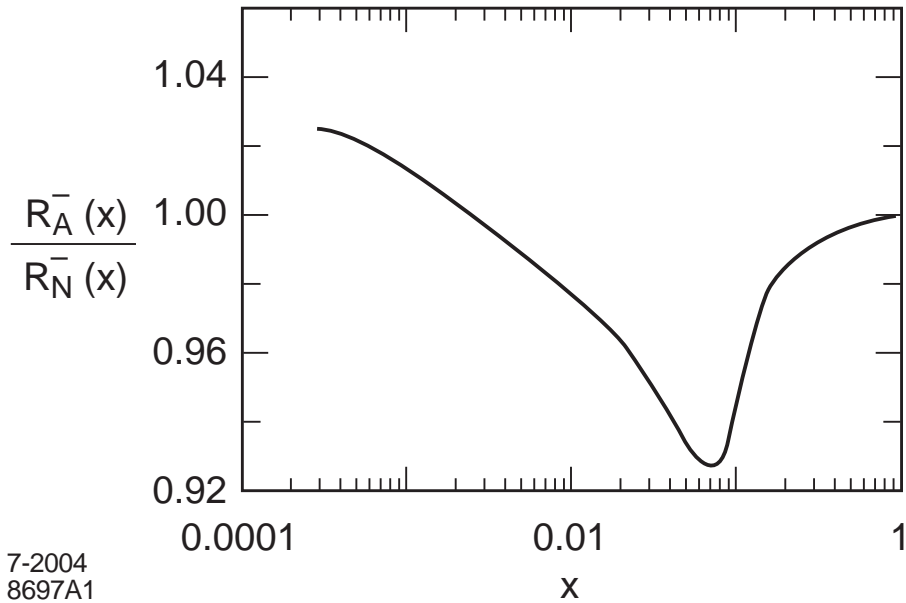}}
\end{center}
\caption[*]{\baselineskip 13pt The  nuclear effect on the
Paschos-Wolfenstein ratio of differential cross sections
$R^-_A(x)/R^-_N(x)$, at $Q^2 = 20~\rm{GeV}^2$.
 \label{8697A01}}
\end{figure}

An alternative way of assessing the nuclear corrections is through
a modified Paschos-Wolfenstein ratio $R^-_A(x)/R^-_N(x)$, in which
instead of the total cross section we consider the corresponding
differential cross sections, as shown in Fig.~\ref{8697A01}. In
this case we have taken $Q^2 = 20~\rm{GeV}^2$, which is the
average value of the NuTeV experiment.

In our numerical analysis we studied the influence of nuclear
effects on the extraction of $\sin^2\theta_W$ from the observable
$R_A^{^-}$, taking into account some kinematical cut-offs specific
to the NuTeV experiment.

The differential cross sections for CC and NC
(anti-)neutrino-nucleus deep inelastic scattering are given
by~\cite{Leader}
\begin{eqnarray}
{\frac{d^2 \sigma_{CC}^{\nu, \bar{\nu} }}{dxdy}}^{(A)} &=&
\frac{G_F^2}{\pi}\ m_N\ E_{\nu, \bar\nu}
\left\{ x y^2 F_1^{W^\pm (A)}(x,Q^2) +
\right. \\ \nonumber
 &+& \left. \left(1-y - \frac {xy m_N}{2 \ E_{\nu, \bar\nu}}\right)
F_2^{W^\pm (A)}(x,Q^2) \pm \left(y-\frac{y^2}{2}
\right) x F_3^{W^\pm (A)} (x,Q^2) \right\},
\end{eqnarray}
for the CC reaction, and
\begin{eqnarray}
{\frac{d^2 \sigma_{NC}^{\nu,\bar{\nu} } }{dxdy} }^{(A)} &=&
\frac{G_F^2}{\pi}\ m_N\ E_{\nu, \bar\nu}\left \{ x y ^2 F_1^{Z (A)}(x,Q^2) +
\right. \\  \nonumber
 &\quad& + \left. \left(1-y - \frac {xy m_N}{2 \ E_{\nu, \bar\nu}}\right)
F_2^{Z (A)}(x,Q^2) \pm \left(y-\frac{y^2}{2} \right) x F_3^{Z (A)} (x,Q^2)
\right \},
\end{eqnarray}
for the NC reaction.

In the event selection, the NuTeV Collaboration applied the cut off
\begin{eqnarray}
\label{Ecal} 20 ~~\rm{GeV} \leq E_{\rm cal} \leq 180 ~~\rm{GeV},
\end{eqnarray}
for a visible energy deposit to the calorimeter $E_{\rm cal}$. The
lower limit ensures full efficiency of the trigger, allows for an
accurate vertex determination and reduces cosmic ray background.

Therefore we will calculate the observables $R_A^{\nu(\bar{\nu})}$
and $R_A^{^-}$ imposing the same cut off on the energy $E_h$ of
the final hadronic state $X$ assuming $E_h = E_{\rm cal}$. Since
$E_h \approx \nu,$ we can write the kinematical variables averaged
over the (anti-)neutrino flux as
\begin{equation}
x=\frac{Q^2} {2 M_N E_{\rm cal}} \leq 1,\ \ \ \ \ \ y=\frac
{E_{\rm cal}} {\langle E_{\nu(\bar{\nu})}\rangle} \leq 1.
\end{equation}
For the average energies of the neutrino and antineutrino beams we
take the values $\langle E_\nu \rangle = 120~\rm{GeV}$ and
$\langle E_{\bar{\nu}}\rangle =112~\rm{GeV}$, as in the NuTeV
experiment~\cite{Zeller-private}.

We will assume a modified version of the Paschos-Wolfenstein
relation:
\begin{eqnarray}
\label{RNMOD}
R_N^{^-} (\sin^2\theta_W) &=& \rho_0^2 (1+\varepsilon) \left(\frac{1}{2} -
\sin^2\theta_W\right)\\ \nonumber  &=& \rho^2 \left(\frac{1}{2} -
\sin^2\theta_W\right),
\end{eqnarray}
where, $\rho^2=\rho_0^2 (1+\varepsilon)$ with a modified factor
$(1+\varepsilon)$ due to strange quark, isospin breaking,
threshold corrections for heavy quarks production, and so on. We
further assume that the Paschos-Wolfenstein relation  can be
applied to the scattering on  a nuclear target $A$,
\begin{eqnarray}
R_A^{^-}(\sin^2\theta_W) = \rho^2 \left(\frac{1}{2} -
(\sin^2\theta_W +\Delta \sin^2\theta_W) \right). \label{RAMOD}
\end{eqnarray}
with a correction $\Delta \sin^2\theta_W$ to the weak-mixing
angle. In Fig.~\ref{bsy1f10}(a), we show the $\sin^2\theta_W$
dependence in the ratio $R_{A/N}^-$. We estimate $\Delta
\sin^2\theta_W$ in the following way. First, we use the cross
sections to calculate the Paschos-Wolfenstein ratios
$R_A^{^-}(\sin^2\theta_W)$ and $R_N^{^-}(\sin^2\theta_W)$ with
various values of $\sin^2\theta_W$. Second, we extract $\rho^2$ by
means of Eq.~(\ref{RNMOD}). In principle, $\rho^2$ should be
different for various values of $\sin^2\theta_W$.  We find a  weak
dependence of $\rho^2$ on $\sin^2\theta_W$ and $\rho^2\simeq
1.04$. Finally, we use the obtained $\rho^2$ to extract the
shadowing/antishadowing effect on the weak-mixing angle $\Delta
\sin^2\theta_W$ from Eq.~(\ref{RAMOD}). The results are given in
Fig.~\ref{bsy1f10}(b).

\begin{figure}[htb]
\begin{center}
\leavevmode {\epsfysize=6cm \epsffile{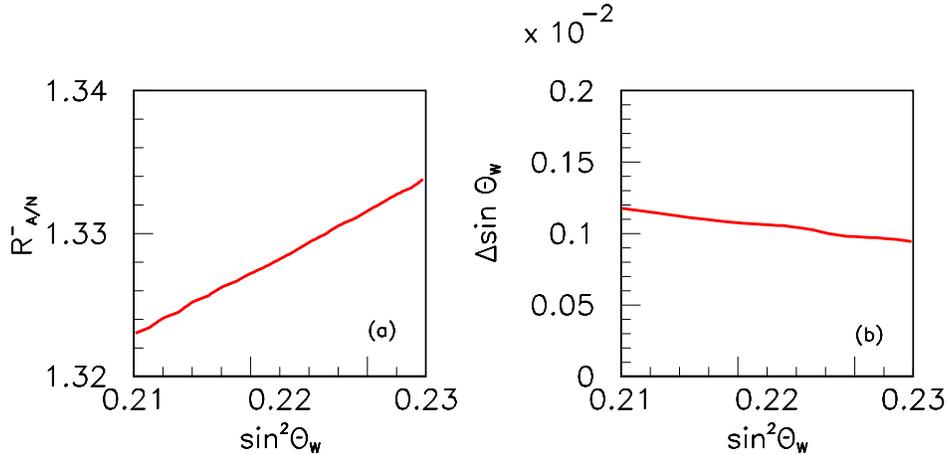}} 
\end{center}
\caption[*]{\baselineskip 13pt
 (a) The  $\sin^2\theta_W$ dependence in $R_{A/N}^-$; (b) The
nuclear shadowing/antishadowing corrections to the  $\sin^2\theta_W$.
 \label{bsy1f10}}
\end{figure}

We have performed a numerical calculation at the average $\langle
Q^2 \rangle=20~\rm{GeV}^2$ of the NuTeV experiment and have found
that the modification to the weak-mixing angle is approximately
$\delta \sin^2\theta_W=0.001.$  The value of $\sin^2\theta_W$
determined from the NuTeV experiment, without including nuclear
shadowing/antishadowing due to multiple scattering, is in absolute
value $ 0.005$  larger than the best value obtain from other
experiments. The model used here to compute nuclear
shadowing/antishadowing effect would reduce the discrepancy
between the  neutrino and electromagnetic measurements of
$\sin^2\theta_W$ by about $20\% $. Together with the charge
symmetry  violation contributions to the neutrino
reactions~\cite{LT03},  about half of the difference between the
standard model and  the NuTeV result can be accounted for.
We also note that the antishadowing effects we predict are
most important in the antineutrino data, which is less sensitive to
$\sin^2\theta_W .$

\section{Conclusions}

We have investigated nuclear shadowing and antishadowing effects
arising from the multiple scattering of quarks and anti-quarks in
the nucleus.   The effective quark-nucleon scattering amplitude
includes Pomeron and Odderon contributions from multi-gluon
exchange as well as Reggeon quark-exchange contributions. The
model is constrained by measurements of the nuclear structure
functions in deep inelastic electron and muon scattering as well
as the Regge behavior of the non-singlet structure functions. We
have also noted the possibility of obtaining an antishadowing
contribution from one-gluon exchange $\times$ Reggeon exchange,
assuming the existence of hidden-color components in the nuclear
wavefunction. We have shown that the coherence of these
multiscattering nuclear processes leads to shadowing and
antishadowing of the electromagnetic nuclear structure functions
in agreement with measurements. The momentum sum rule is not
satisfied in a nuclear target by balancing the shadowing and
antishadowing of the leading-twist nuclear quark distributions;
however the momentum sum rule can still be satisfied if there is a
compensating change in the nuclear gluon distribution.

Our analysis leads to substantially different nuclear
antishadowing for charged and neutral current reactions; in fact,
the neutrino and antineutrino DIS cross sections are each modified
in different ways due to the various allowed Regge exchanges. The
non-universality of nuclear effects will modify the extraction of
the weak-mixing angle $\sin^2\theta_W$, particularly because of
the strong nuclear effects for the $F_3$ structure function. The
shadowing and antishadowing of the strange quark structure
function in the nucleus can also be considerably different than
that of the light quarks. We thus find that part of the anomalous
NuTeV result for $\sin^2\theta_W$ could be due to the
nonuniversality  of nuclear antishadowing for charged and neutral
currents. Our picture also implies non-universality for the
nuclear modifications of spin-dependent structure functions.

We have found in our analysis that the antishadowing of nuclear
structure functions depends in detail on quark flavor. Careful
measurements of the nuclear dependence of charged, neutral, and
electromagnetic DIS processes are thus needed to establish the
distinctive phenomenology of shadowing and antishadowing and to
make the NuTeV results definitive. It is also important to map out
the shadowing and antishadowing of each quark component of the
nuclear structure functions to illuminate the underlying QCD
mechanisms. Such studies can be carried out in semi-inclusive deep
inelastic scattering for the electromagnetic current at Hermes and
at Jefferson Laboratory by tagging the flavor of the current quark
or by using pion and kaon-induced Drell-Yan reactions. A new
determination of $\sin^2\theta_W$ is also expected from the
neutrino scattering experiment NOMAD at CERN~\cite{Petti}. A
systematic program of measurements of the nuclear effects in
charged and neutral current reactions could also be carried out in
high energy electron-nucleus colliders such as HERA and eRHIC, or
by using high intensity neutrino beams~\cite{Geer}.

\vspace{0.3cm}

\appendix{\large{\bf Appendix: A non-abelian source for antishadowing}}

\vspace{0.3cm}
\begin{figure}[htbp]
\begin{center}
\leavevmode {\epsfysize=10cm \epsffile{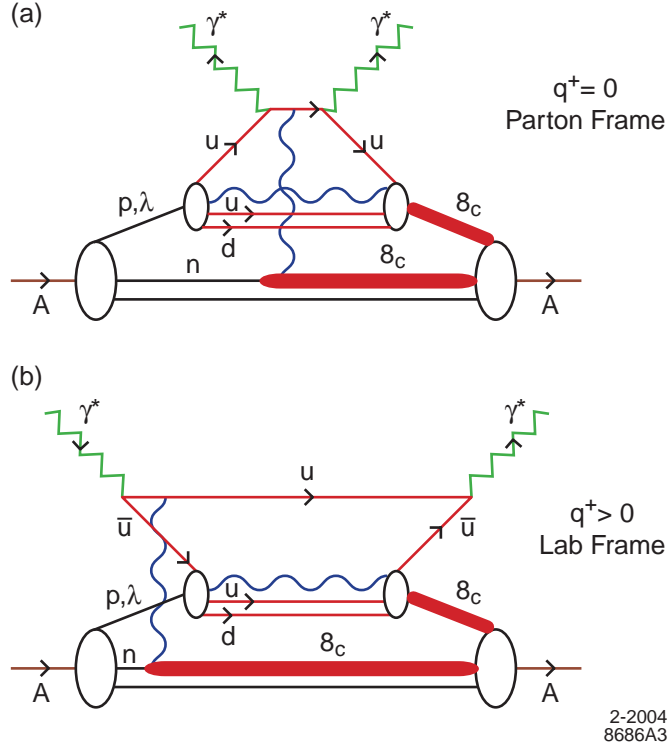}}
\end{center}
\caption[*]{\baselineskip 13pt Representation of leading-twist QCD
``hidden color" contributions to the nuclear structure function
from the absorptive part of the forward virtual Compton amplitude
{\it Im}$~ T(\gamma^* A \to \gamma^* A).$ (a) Illustration of a
two-step contribution in the $q^+ =0, q^2_\perp = Q^2$ parton
model frame---deep inelastic lepton scattering on a valence quark
of a target proton followed by the final-state single-gluon
interaction of the outgoing quark on a target neutron. The proton
and neutron are both color excited to color-octet states. The
amplitude requires the presence of hidden-color components in the
nuclear wavefunction. (b) Illustration of the physics of the
two-step process shown in (a), but in the laboratory frame where
$q^+ > 0$.  The $u \bar u$ fluctuation of the virtual photon first
scatters via  a single gluon exchange on a neutron; this is then
followed by the annihilation of the $\bar u$ quark on a proton.
The proton and neutron are both color excited to color-octet
states.  The amplitude requires the presence of hidden-color
components in the nuclear wavefunction. The two-step amplitude can
interfere constructively with the Regge-behaved single-step
annihilation amplitude on the proton alone, thus producing
antishadowing. Similar processes occur in the case of the weak
currents. \label{8686A03}}
\end{figure}

We can identify a further antishadowing contribution specific to
the non-abelian theory. Consider once again Fig.~\ref{8686A01} for
$\gamma^* A \to \gamma^* A$, but replace the two exchanged gluons
with just a single gluon (see Fig.~\ref{8686A03}). For simplicity
we display the case of a deuteron target. In Fig.~4(a) the
exchanged gluon attaches to the struck $u$ quark valence
constituent  of the proton at the top of the diagram changing its
color.  This also changes the scattered proton $p^\prime$ to a
color octet. The exchanged gluon also transforms the spectator
neutron into a color octet. Thus if the deuteron wavefunction
contains hidden color  $ | 8_C 8_C\rangle $ components, this
process interferes with the one step diagram with no final state
interactions.

The deuteron certainly has hidden-color components---one only has
to exchange a gluon between the nucleons in the deuteron
LFWF~\cite{Brodsky:1983vf}. The large magnitude of the deuteron
form factor also demands hidden color components (\cite{Farrar}).
The calculation of the one-gluon exchange effects is very similar
to our Odderon analysis. The one-gluon exchange amplitude behaves
as $s^1$ and a nearly real phase. Like the Odderon,  it has $C=-$
and couples with opposite sign to the  $q$ and $\bar q.$ A
complication is how to understand the Reggeon exchange amplitude
on the proton since the $u$ and $\bar u$ in the $t$ channel now
are in a color octet configuration. Nevertheless, the Odderon
calculation serves as a model for the one-gluon exchange
contribution as well and its effect on antishadowing.

{\bf Acknowledgments: } We thank Stefan Kretzer, Kevin McFarlane, and Paul Hoyer
for helpful comments, and Alfonso Zerwekh for assistance with the
numerical analysis.

\end{document}